\documentclass[sigconf,nonacm]{acmart}

\AtBeginDocument{%
  \providecommand\BibTeX{{%
    \normalfont B\kern-0.5em{\scshape i\kern-0.25em b}\kern-0.8em\TeX}}}

\setcopyright{acmcopyright}
\copyrightyear{2020}
\acmYear{2020}
\acmDOI{10.1145/1122445.1122456}

\acmConference[FODS-2020]{ACM-IMS Foundations of Data Science Conference}{October 18--20, 2020}{Seattle, WA}
\acmBooktitle{FODS-2020: ACM-IMS Foundations of Data Science Conference,
  October 18--20, 2018, Seattle, WA}
\acmPrice{15.00}
\acmISBN{978-1-4503-XXXX-X/18/06}

\usepackage{bbm}
\usepackage{xcolor,bm,amsmath,graphicx,comment,bigints}
\usepackage{algorithm,algorithmic}
\graphicspath{{images-similar/}}

\definecolor{plum}{rgb}{0.45,0,.66}
\definecolor{mythistle}{rgb}{.99,.195,.133}
\definecolor{myred}{cmyk}{0.000000,1.000000,1.000000,0.1}
\definecolor{myblue}{cmyk}{1.000000,0.750000,0.000000,0.1}
\definecolor{mybgn}{cmyk}{0.850000,0.350000,0.000000,0.1}
\definecolor{mygrn}{cmyk}{0.750000,0.000000,1.000000,0.2}

\newcommand{\bl}{\begin{itemize}}
\newcommand{\el}{\end{itemize}}
\newcommand{\be}{\begin{enumerate}}
\newcommand{\ee}{\end{enumerate}}

\newcommand{\bea}{\begin{eqnarray*}}
\newcommand{\eea}{\end{eqnarray*}}

\newcommand{\beq}{\begin{equation}}
\newcommand{\eeq}{\end{equation}}

\newcommand{\bmx}{\left[ \begin{array}}
\newcommand{\emx}{\end{array} \right]}

\begin{document}

\title{Using Wavelets to Analyze Similarities in Image-Classification Datasets}

\author{Roozbeh Yousefzadeh}
\affiliation{
\institution{Yale University, New Haven, CT}
}
\email{roozbeh.yousefzadeh@yale.edu}

%
%
%
%
%
%

\renewcommand{\shortauthors}{Yousefzadeh}

\begin{abstract}
Deep learning image classifiers usually rely on huge training sets and their training process can be described as learning the similarities and differences among training images. But, images in large training sets are not usually studied from this perspective and fine-level similarities and differences among images is usually overlooked. This is due to lack of fast and efficient computational methods to analyze the contents of these datasets. Some studies aim to identify the influential and redundant training images, but such methods require a model that is already trained on the entire training set. Here, using image processing and numerical analysis tools we develop a practical and fast method to analyze the similarities in image classification datasets. We show that such analysis can provide valuable insights about the datasets and the classification task at hand, prior to training a model. Our method uses wavelet decomposition of images and other numerical analysis tools, with no need for a pre-trained model. Interestingly, the results we obtain corroborate the previous results in the literature that analyzed the similarities using pre-trained CNNs. We show that similar images in standard datasets (such as CIFAR) can be identified in a few seconds, a significant speed-up compared to alternative methods in the literature. By removing the computational speed obstacle, it becomes practical to gain new insights about the contents of datasets and the models trained on them. We show that similarities between training and testing images may provide insights about the generalization of models. Finally, we investigate the similarities between images in relation to decision boundaries of a trained model.
\end{abstract}



\maketitle

\section{Introduction}

Studying the similarities and differences among images in training sets may provide valuable insights about the data and the models trained on them. For example, we may identify redundancies and/or anomalies in the training sets, or we may gain insights about the generalization of models on testing sets and understand their misclassifications in relation to training sets. Some studies in the literature aim to identify redundant and influential images in datasets. However, such analysis is performed in a post-hoc way and require a model that is trained on all the data. Here, we leverage image processing and numerical analysis methods for analyzing image classification datasets prior to training. Our computational approach is very fast compared to the methods in the literature. Interestingly, the redundant images that we identify are the same as the results of previous methods that require pre-trained models, confirming the validity of our approach to use wavelets to analyze contents of image classification datasets.

The fast computation makes it practical to provide many useful insights about the contents of image classification datasets. We show that a similarity matrix and analysis of eigen-gaps of a graph Laplacian can provide an estimate about portion of redundancies in the datasets, prior to training a model, which is useful for many applications that rely on automated gathering of vast amount of images with lots of redundancies. Our method can also be used to identify influential images and to create a graphical model representing the contents of image classification datasets.

Moreover, analyzing the similarities across training and testing sets may provide insights about the generalization in deep learning.

\begin{figure}[h]
\begin{center}
   \fbox{\includegraphics[width=0.2\linewidth]{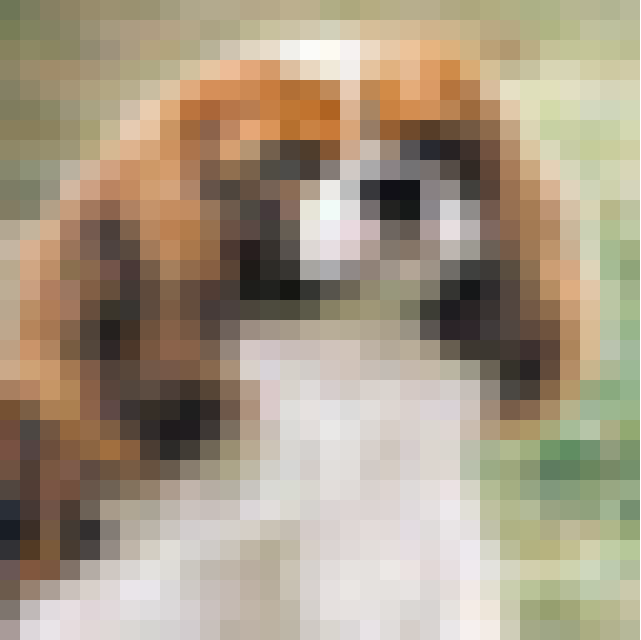}
   \includegraphics[width=0.2\linewidth]{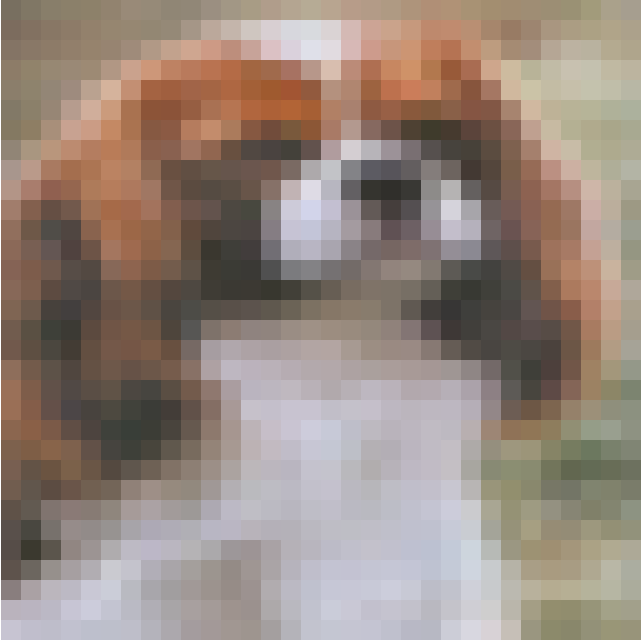}
   \includegraphics[width=0.2\linewidth]{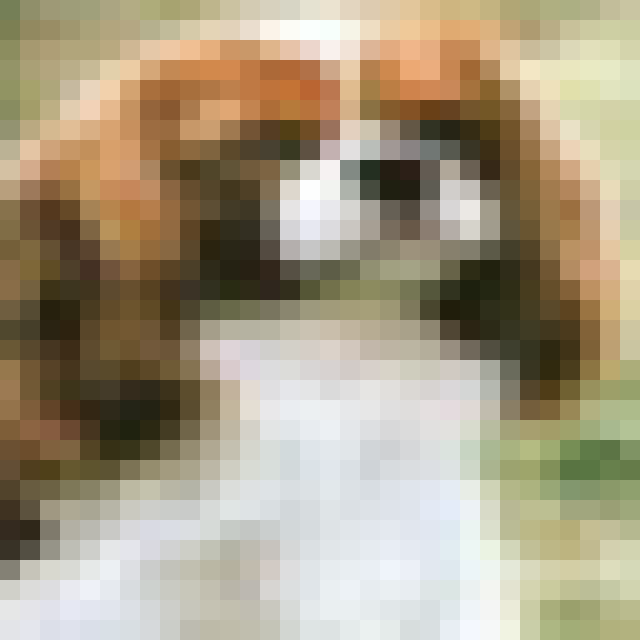}}\\
   \fbox{\includegraphics[width=0.2\linewidth]{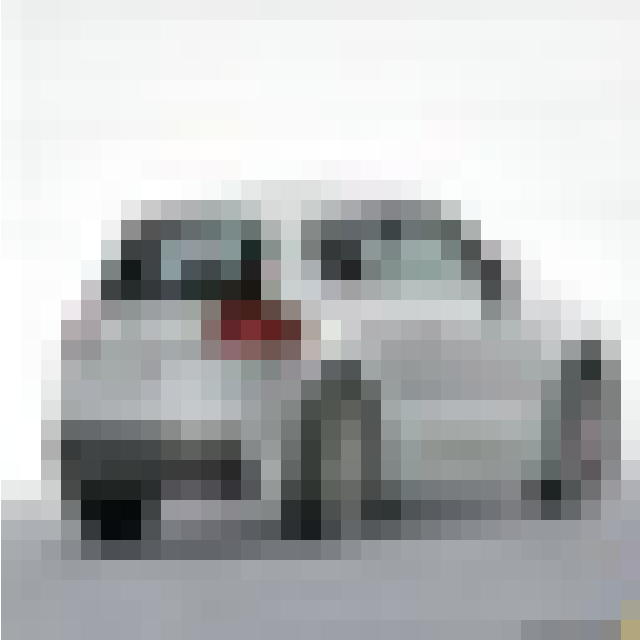}
   \includegraphics[width=0.2\linewidth]{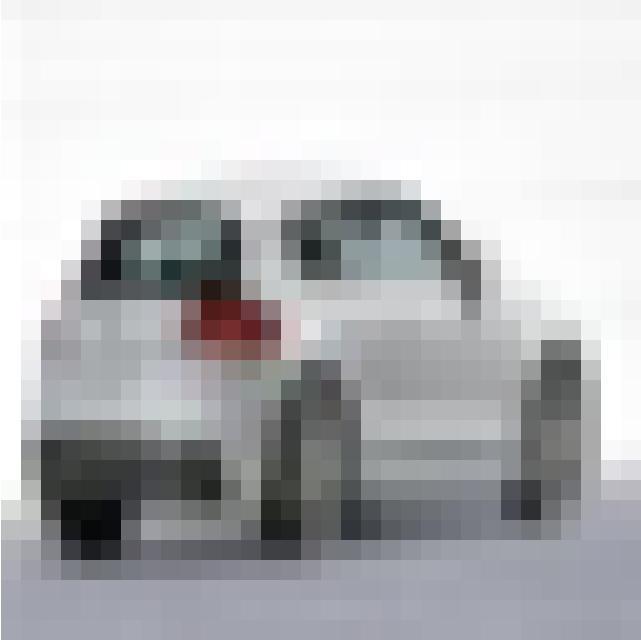}
   \includegraphics[width=0.2\linewidth]{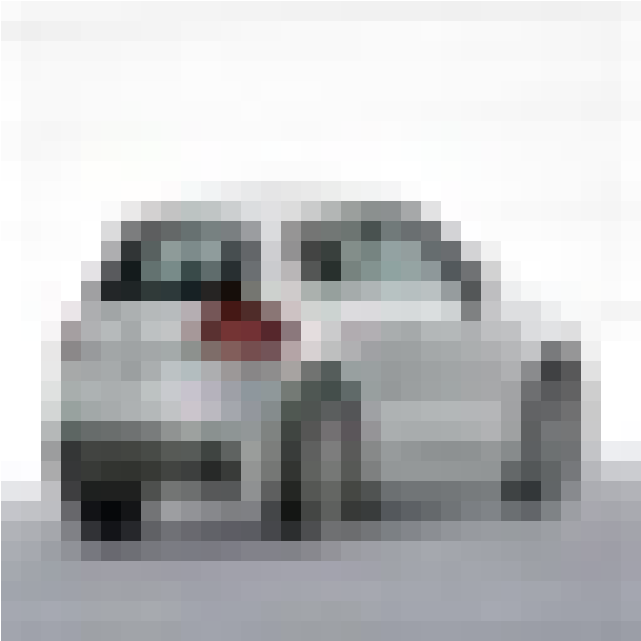}}\\
\end{center}
   \caption{Example of nearly identical images in CIFAR-10 training set.}
\label{fig_cifar10_redund}
\end{figure}

\vspace{-.5cm}

\begin{figure}[h]
\begin{center}
\fbox{\includegraphics[width=0.2\linewidth]{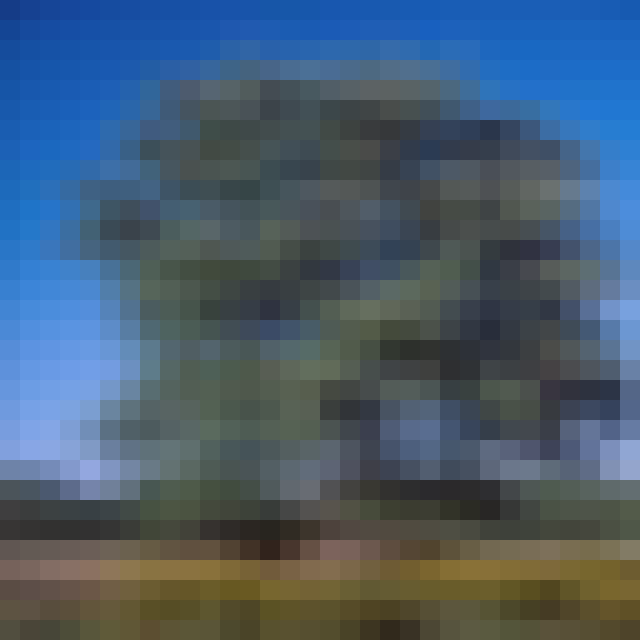}
   \includegraphics[width=0.2\linewidth]{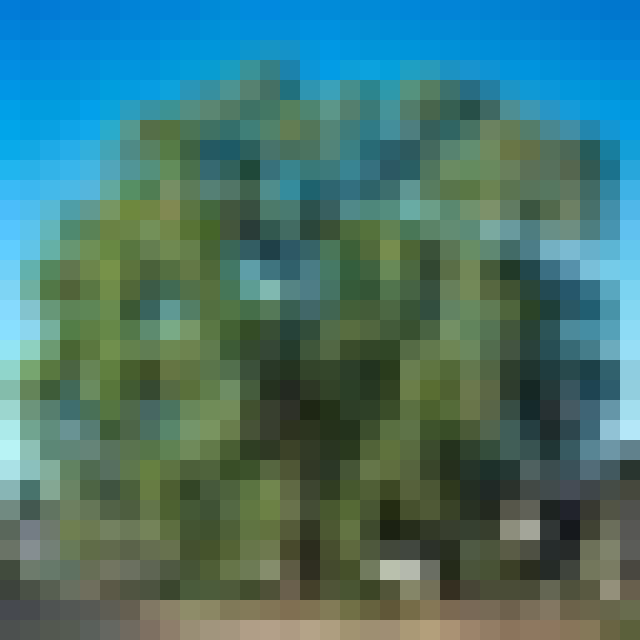}}
\fbox{\includegraphics[width=0.2\linewidth]{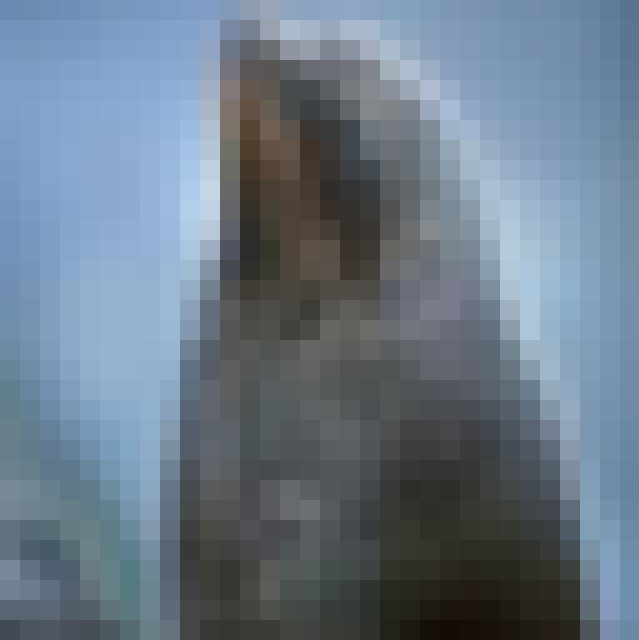}
   \includegraphics[width=0.2\linewidth]{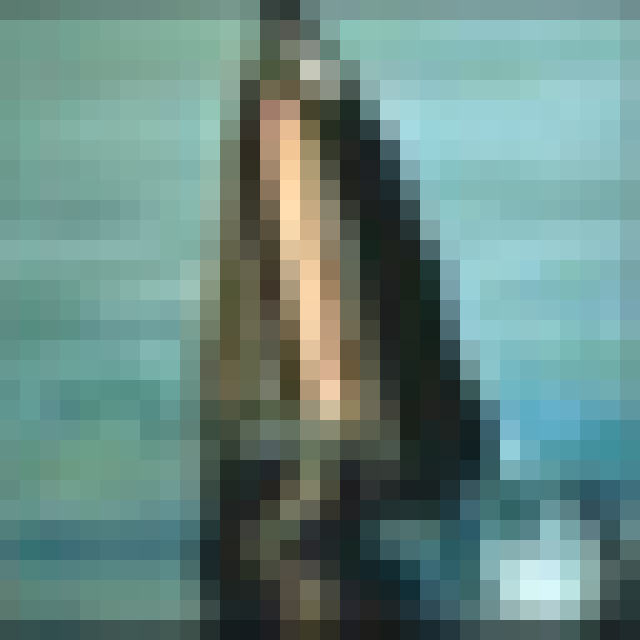}}
\end{center}
   \caption{Example of similar images with different labels in CIFAR-100 training set: oak and maple tree (left), whale and seal (right).}
\label{fig_cifar100_inf}
\end{figure}

Figure~\ref{fig_cifar10_redund} shows examples of redundancies present in the CIFAR-10 dataset and Figure~\ref{fig_cifar100_inf} shows examples of similar images with different labels in the CIFAR-100 dataset. Our method identifes all of such similarities in a few seconds, with no need for a trained model.

\subsection{Image classifiers and their decision boundaries}

Any classification model is defined by its decision boundaries, and hence, training process of a model can be viewed as defining those decision boundaries for the model. Consider for example, the case of training a linear regression model. What happens during the training is basically defining the location of the regression line (i.e., decision boundary) to partition the input space. Training of an image classification model is also partitioning its input space, although such partitions can be geometrically complex in the high dimensional space.

\nocite{fawzi2018empirical}

From this perspective, any training image that does not affect the decision boundaries of a model can be considered redundant, and any training image that affects the partitioning of the input space can be considered influential. 
Therefore, when a group of images from the same class are similar, it is likely that only one of them would suffice to define the necessary decision boundary in that neighborhood in the domain. 

On the other hand, when images of different class are similar (e.g., see Figure~\ref{fig_cifar100_inf}), they would be influential, because they are similar, and learning them would cause the model to define the necessary decision boundaries between them. We study these conjectures in our numerical experiments.\footnote{As in most machine learning tasks, the underlying assumption is that the data in the testing set generally comes from a similar distribution as the training set and learning the similarities and differences among images in the training set is the key to achieving good generalization.}


\subsection{Our plan}

We propose one general approach and two specific implementations to analyze the similarities in image-classification datasets. The general approach is to use wavelets to measure the similarities among images and to analyze those similarities to provide insights about the contents of datasets.

Our first implementation (Algorithm~\ref{alg_small}) is fast and effective as it decomposes images and clusters them based on their similarities. Any clustering method can be used. Our second implementation (Algorithm~\ref{alg_spectral}) performs a more thorough analysis of datasets by forming a similarity matrix and investigating the eigen-gaps of a graph Laplacian. We later show that this can provide a graphical model and reveal possible anomalies in the dataset.

We show the effectiveness of our methods on standard object recognition datasets such as CIFAR-10 and CIFAR-100 \citep{krizhevsky2009learning}, and also one class of Google Landmarks dataset \citep{noh2017large}.


In Section~\ref{sec:literature}, we review the related work. In Sections~\ref{sec:method1} and \ref{sec:method2}, we describe each of our methods, respectively. Section~\ref{sec:results} includes our numerical results, and finally, in Section~\ref{sec:conclusion}, we discuss our conclusions and directions for future research.

\section{Related work} \label{sec:literature}

One of the early studies in the literature is by \citet{ohno1998improving} which reports existence of redundancies in medical training sets, but their computational method is not practical for modern applications of deep learning. 

There are studies that measure the influence of training points based on the derivatives of the loss function of a trained model. \citet{guo2008discriminative} studied the derivatives of mini-batches to select subsets of unlabeled data but their method is specific to active learning, where there is a stream of new training data. \citet{vodrahalli2018all} showed that choosing training images with most diversity of derivatives can speed up the training, but their analysis is based on models trained on all the data. Recently, \cite{sankararaman2019impact} explained the speed of training in terms of correlation between gradients.

There are also methods that consider the value of loss function or norm of gradients \citep{loshchilov2015online,alain2015variance,katharopoulos2018not}. Such measures are useful to speed up the training, but not necessarily meaningful proxies to compare similarity of images. Additionally, since they require computation of loss and gradient of a model, they are more expensive than direct comparison of images using image processing tools.

\citet{lapedriza2013all} proposed a greedy method to sort the data points in training sets based on their importance for training. For the sorting, they define a ``training value" which requires the model to be separately trained from scratch for each training image.

Influence functions have been used to quantify the effect of individual training data on a trained model \citep{koh2017understanding,koh2019accuracy}, but the assessment of influence requires a model trained on all the data.

\citet{carlini2018prototypical} developed a method that first projects the images from the pixel space into a two dimensional space, and then performs clustering in the low dimensional space, but the projection requires a model trained on all the training set.


There are other studies focused on interpretability of image classification models that make use of prototypes, for example, \cite{kim2014bayesian,li2017deep,chen2018looks}. Such methods aim to train a model such that its output is explainable in terms of similarity to prototypes. To compare an image with a prototype, \citet{chen2018looks} inverts the $\ell_2$ norm distance between the output of a trained convolutional layer and the prototype.

\citet{meletis2019data} used a Gaussian Mixture Model (GMM) to identify visually similar images using a pre-trained model.

\citet{birodkar2019semantic} used clustering of images in a semantic space to identify redundant images. The semantic space in their study is the intermediate output of a trained model. \citet{barz2019we} also showed that redundant images in CIFAR datasets can be identified using the output of an average pooling layer.


\citet{chitta2019less} showed that a portion of some training sets can be removed leading to no loss of accuracy. Their method trains an ensemble of deep neural networks on all the training set in order to identify such redundancies. \citet{yousefzadeh2019interpreting} showed that the distance of training images to the decision boundaries of a trained model can identify the most influential training data.

All the approaches above identify the influence of training images through the lens of a model that is trained on all the data. Here, we show that using wavelets, one can analyze the images and obtain valuable information about them, before engaging in the training process.

\citet{achille2019task2vec} studied the similarities between different image classification tasks, e.g., classifying different kinds of birds vs different kinds of mammals, which can provide valuable information about the nature of those classification tasks, however, their approach requires model training.

Finally, we note that there are unsupervised methods that aim to create an embedding for groups of images. For example, \cite{vo2019unsupervised} solves an optimization problem for each image pair to measure their similarity and then creates an embedding based on that information. However, such methods are not scalable to analyze an entire training set.

\nocite{yousefzadeh2019thesis}

\section{Finding similar (redundant and influential) images in the data}  \label{sec:method1}

Here, we develop a simple and fast algorithm to identify similar images in training sets. Our algorithm first decompose the images using wavelets, then chooses a subset of wavelet coefficients that have the most variation among images. Finally, it clusters the images based on their wavelet coefficients. Images that are similar will appear in same clusters and that lead to identifying influential and redundant images. Repeating the same analysis on testing sets can also provide insights about testing sets.


\subsection{Wavelet decomposition of images}





We use the wavelet transformation of images as a mean to identify similar training data. Wavelets are a class of functions that have shown to be very effective in analyzing different kinds of data, especially images and signals. Wavelets can also be used to analyze functions and operators \citep{daubechies1992ten}. In both image processing and signal processing, wavelets have been effective in compressing the data and also in identifying actual data from the noise \citep{chui2016introduction}.

The main idea here is to decompose each image into different frequency components and then analyze the components among the images to identify their similarities and differences. Wavelets allow us to analyze images at different resolutions and therefore enable us to compare them effectively.

Decomposing an image using a wavelet basis is basically convolving the wavelet basis over the image. This is similar to the operation performed by convolutional neural networks, as CNNs also convolve a stencil with the input image. Therefore, our method of analyzing and comparing images is similar to the computational method that will be used by the classification models. In our results, we see that wavelets identify the similarity of images, in the same way that a pre-trained ResNet-50 does.

In this paper we use wavelets, but we note that shearlets \citep{kutyniok2012shearlets} are also a class of functions with great success in analyzing images, and therefore they can be considered instead of wavelets.

\subsection{Extracting a subset of influential wavelet coefficients}

We are interested in the similarities and differences among subgroups of images. But, as we will show in our numerical experiments, many of the wavelet coefficients can be similar among all images in a training set and therefore, not helpful for our analysis.

When computing wavelet decomposition, one can use different resolutions to convolve the wavelet basis with images. Since we are concerned with the overall similarity of images, there will be no need to extract the wavelet coefficients on a very fine level. Therefore, even for relatively large images in Imagenet and Google Landmarks datasets, one can extract a relatively small number of wavelet coefficients by convolving the images with high pass filters.

Once we have computed the wavelet coefficients for images, we use rank-revealing QR factorization \citep{chan1987rank} to choose a subset of coefficients that are most linearly independent among the images. Rank-revealing QR algorithm and also its variation, pivoted QR algorithm \citep{golub2012matrix} decompose a matrix by computing a column permutation and a QR factorization of a given matrix. The permutation matrix orders the columns of the matrix such that the most linearly independent (non-redundant) columns are moved to the left. The rows of our matrix represent the images and its columns represent wavelet coefficients.

The obtained permutation matrix allows us to choose a subset of most independent columns. This dimensionality reduction in the wavelet space, previously used by \citet{yousefzadeh2019investigating,yousefzadeh2019refining}, can make our next computational step (i.e., clustering) faster.

As an example, consider the 60,000 images in the training set of MNIST dataset. Each image has 784 pixels, leading to 784 wavelet coefficients using the Haar wavelet basis. 32 of those wavelet coefficients are 0 for all images. After discarding those coefficients, the condition number of the training set is greater than $10^{22}$, implying linear dependency of columns. After performing rank-revealing QR factorization, we observe that dropping the last 200 columns in the permutation matrix will decrease the condition number to about $10^6$. All the discarded features will be completely unhelpful in identifying influential and redundant data, because they are either uniform or (almost) linearly dependent among all images.




The cost of computing the rank-revealing QR factorization is $\mathcal{O}(nd^2)$ given $n$ images with $d$ wavelet coefficients \citep{golub2012matrix}. We would not need to compute the entire decomposition as the computation can be stopped as soon as a diagonal element of $R$ becomes small enough compared to its first diagonal element.

We note that the possible computational advantage of using rank-revealing QR depends on the relative values of $n$, $d$, and also the computational cost of clustering method which will be discussed in the following.

%
%
%
%

\subsection{Clustering images based on their wavelet coefficients}

Any clustering method can be utilized to cluster the images based on their wavelet coefficients. The computational effort for clustering depends on the number of observations (size of training set) and also the number of features considered for each image. We suggest performing the clustering on the entire dataset (when possible), noting that it will be a more expensive computation compared to clustering images of each class, separately. Clustering per class would cost less, but would not provide the additional insight about influential images.

The usual trade-offs among the clustering methods apply here as well. Some clustering methods identify the appropriate number of clusters in the data during the clustering, for example Newman's community structure algorithms \cite{newman2004fast}, which would be useful when we are not aware of the percentage of similarities in the dataset. But, when the number of clusters is known beforehand, a less costly clustering method could suffice, for example k-means \cite{lloyd1982least,arthur2007k}.


\subsection{Our algorithm based on wavelet coefficients and clustering}

\begin{algorithm}[h]
\caption{Algorithm for finding similar training images using wavelet coefficients and clustering}
\label{alg_small}
\textbf{Inputs}: Training set $\mathcal{D}^{tr}$, $\tau$, clustering method $\mathcal{M}$, $n_c$  \\
\textbf{Outputs}: Reduced training set $\mathcal{\hat{D}}^{tr}$ and the list of most influential images $\mathcal{I}$\\
\begin{algorithmic}[1] 
%
\STATE Count number of images in $\mathcal{D}^{tr}$ as $n$
\FOR {$i = 1$ to $n$}
	\STATE Compute wavelet coefficients of image $i$, vectorize them and save them in row $i$ of matrix $W$
\ENDFOR
\IF{$n >$ number of wavelet coefficients per image}
	\STATE $[Q,R,P] = \text{RR-QR}(W)$, i.e. perform rank-revealing QR on $W$
	\STATE Choose $m$ as large as possible such that the first $m$ columns of the matrix $WP$ has condition number less than $\tau$
	\STATE Extract the first $m$ columns of $WP$ and save it as $\hat{W}$
\ENDIF
\STATE Perform clustering on $\hat{W}$ using method $\mathcal{M}$ (with $n_c$ clusters)
\FOR {$i = 1$ to $n_c$}
	\IF{there are more than one image in cluster $i$}
		\IF{all images in the cluster are from the same class}
			\STATE Keep the image closest to the center of that cluster and discard other images
		\ELSE
			\STATE Add the images in the cluster to the list $\mathcal{I}$
		\ENDIF
	\ENDIF
\ENDFOR
\STATE Put together remaining images in clusters as $\mathcal{\hat{D}}^{tr}$
\STATE \textbf{return} $\mathcal{\hat{D}}^{tr}$ and $\mathcal{I}$
\end{algorithmic}
\end{algorithm}

Algorithm \ref{alg_small} formalizes the above process in detail. The algorithm first computes wavelet decomposition of all images in the training set and forms them in a matrix $W$, where rows are samples and columns are wavelet coefficients (lines 1 through 4). The next step in the algorithm is to compute the rank-revealing QR factorization of $W$ (line~5). This factorization computes an orthogonal matrix $Q$, an upper-triangular matrix $R$, and a permutation matrix $P$, such that $${W}{P} = {Q} {R}.$$

Algorithm \ref{alg_small} then chooses a subset of $m$ most independent wavelet coefficients according to the permutation matrix (lines 6 and 7). The condition for choosing~$m$ is to maximize its value such that the condition number of the first $m$ columns of $WP$ is less than~$\tau$. For a given dataset, this condition will yield a unique value for~$m$. The best value for $\tau$ could vary based on the properties of the dataset. The trade-off here is that choosing a small $\tau$ will yield a small $m$, making the clustering computation less expensive, while using a very small $m$ may not be able to adequately capture the variations among images and lead to poor results. In our numerical experiments, we found $\tau = 10^5$ to be a good choice. The final stage of the algorithm is to perform clustering, discard the redundant images, and return the list of influential images (lines 9~through~13).


About the number of clusters, $n_c$, its best value would depend on the portion of similar images in a training set which is likely to be unknown. In typical image classification datasets, the portion of similar images (including both same and differing classes) make up less than half of datasets. So, the clustering step can be considered \textit{coarse-graining}. There are many methods available to choose an appropriate value for $n_c$. Later, we describe and use a method that chooses the $n_c$ based on the eigen-gaps of graph Laplacian derived from a similarity matrix.

\section{Comparing images using specialized image processing tools and analyzing a similarity matrix}  \label{sec:method2}

Algorithm~\ref{alg_small} computes the wavelet decomposition of images and compares the images based on the similarity of their wavelet coefficients in the Euclidean space. We show in our results that this is adequate for analyzing the similarity of images in standard datasets for object recognition. However, we note that there are more sophisticated methods to compare images which we consider in this section.

\subsection{Specialized wavelet-based similarity measures between images}

There are many methods in the image processing literature for measuring the similarities between images, for example, \cite{reisenhofer2018haar,wang2005translation,sampat2009complex}. \citet{albanesi2018new} recently proposed a class of metrics to measure the similarity between pairs of images.  Here, we use a relatively recent and widely used method known as the Structural Similarity Index (SSIM), developed by \citet{wang2004image}, which compares images based on local patterns of pixel intensities that have been normalized for luminance and contrast. 

Some of these measures are designed to measure specific kinds of similarity, for example, structural similarity, perceptual similarity, textural similarity, etc. Considering the structure of images and patterns of pixel intensities make the SSIM particularly useful for image classification of objects such as the ones in CIFAR and Imagenet datasets.

We note that the similarity measure should be chosen based on the classification task at hand. For example, in classifying images of skin cancer \citep{tschandl2018ham10000}, the textures present in images may be more influential in classifications, instead of the structure of images. In such case, a texture-based similarity measure such as \citep{zujovic2013structural} may be more effective than the SSIM.

Additionally, many of these measures are tunable. For example, SSIM measures the similarity of images based on three components: luminance, contrast, and structure, and returns an overall similarity score based on them. The weight of components and other tunable parameters of SSIM can be adjusted to measure the specific similarities of interest.

We note that in the image processing literature, image retrieval techniques aim to find images in a dataset that are similar to a base image, some of which use wavelets, for example \cite{loupias2000wavelet}. Our use of wavelet decomposition of images (instead of their pixels) is similar in nature to some of those image retrieval techniques. However, those techniques are not directly applicable for our purpose of analyzing image-classification datasets from the perspective of deep learning. One contribution of our work is to bridge one of the gaps between image processing and deep learning literatures.


%




\subsection{Our algorithm for similarity matrix analysis}

Here, we develop Algorithm~\ref{alg_spectral} to perform the analysis via a similarity matrix.

For each image pair in a training set, we compute their similarity using a function of choice $\mathcal{F}$, and form a similarity matrix, $\mathcal{S}$ (lines 3 through 7 in Algorithm~\ref{alg_spectral}).

As mentioned in previous section, $\mathcal{F}$ should be chosen based on the patterns present in images and the type of similarities and differences among them. In some datasets, we observed that even the cosine similarity between vectors of wavelet coefficients can be insightful.


After computing the $\mathcal{S}$, our algorithm computes the eigenvalues of its graph Laplacian (line 8). Laplacian is the matrix representation of a graph corresponding to the similarity matrix. Instead of computing the precise eigenvalues, one can compute an estimate to the distribution of eigenvalues, using a Lanczos-based method, e.g., the method developed by \citet{dong2019network}. The next step is to choose the number of clusters, $n_c$, based on the number of eigen-gaps of the graph Laplacian, as suggested by \citet{von2007tutorial} (line 9). These two lines of the algorithm can be skipped, if the user wants to use a specific $n_c$, for example based on prior knowledge about the dataset.

The algorithm then completes the spectral clustering. The process of discarding redundant training data is similar to Algorithm~\ref{alg_small}. Overall, this approach has $\mathcal{O}(n^3)$ because of the spectral clustering.

\begin{algorithm}[h!]
\caption{Algorithm for analyzing training images using a similarity matrix}
\label{alg_spectral}
\textbf{Inputs}: Training set $\mathcal{D}^{tr}$, threshold on eigen-gaps $\gamma$, similarity function $\mathcal{F}$  \\
\textbf{Outputs}: Reduced training set $\mathcal{\hat{D}}^{tr}$ and the list of most influential images $\mathcal{I}$\\
\begin{algorithmic}[1] 
%
\STATE Count number of images in $\mathcal{D}^{tr}$ as $n$
\STATE Initialize similarity matrix $S_{n \times n}$ as matrix of zeros
\FOR {$i = 1$ to $n-1$}
	\FOR {$j = i+1$ to $n$}
		\STATE $S_{i,j} = S_{j,i} = \mathcal{F}(\text{image } $i$, \text{image } $j$)$
	\ENDFOR
\ENDFOR
\STATE Compute the eigenvalues of the graph Laplacian of $\mathcal{S}$, or an estimate to the distribution of eigenvalues.
\STATE Count the eigen-gaps larger than $\gamma$ and use it as $n_c$
\STATE Complete the spectral clustering on $\mathcal{S}$ with $n_c$
\FOR {$i = 1$ to $n_c$}
	\IF{there are more than one image in cluster $i$}
		\IF{all images in the cluster are from the same class}
			\STATE Keep the image closest to the center of that cluster and discard other images
		\ELSE
			\STATE Add the images in the cluster to the list $\mathcal{I}$
		\ENDIF
	\ENDIF
\ENDFOR
\STATE Put together remaining images in clusters as $\mathcal{\hat{D}}^{tr}$
\STATE \textbf{return} $\mathcal{\hat{D}}^{tr}$ and $\mathcal{I}$
\end{algorithmic}
\end{algorithm}

A low-cost alternative to spectral clustering is to check for each image, whether there are any other images similar to it and keep only one image from each group of images that have SSIM larger than a threshold. This basically requires investigating individual rows in the upper triangular section of $\mathcal{S}$. The complexity of such algorithm is $\mathcal{O}(n^2)$ which might be appealing for large training sets.

We note that SSIM can be used in conjunction with other clustering methods, for example, k-means, leading to $\mathcal{O}(n)$ cost. In such approach, larger similarity between an image pair will be interpreted as closer distance between them and vice versa.

For datasets with large images, it is possible to compress the images first and then perform the analysis. The effectiveness of such approach would depend on the contents of images in the dataset.

\section{Insights about generalization of image classifiers}

Generalization of image classification models is an open research problem. Deep learning has been impressively successful in image classification, but the reason behind the accuracy of models and also their mistakes is not well understood. \citet{zhang2016understanding} famously showed that many traditional approaches (model properties or regularization techniques) fail to explain the generalization in deep learning and hence initiated a series of fundamental studies about generalization. Some recent studies relate the generalization to decision boundaries of models, for example \cite{marginbased2019,elsayed2018large}. \citet{huang2019understanding} used a visualization method to provide insights about generalization. Kernel methods \cite{belkin2018understand} and compression methods \cite{arora2018stronger,li2020understanding} are used to study generalization, too. From the optimization perspective, \citet{arora2019fine} studied the generalization of models by studying the details of training process. All of these approaches study the trained models and/or training process, and usually overlook the fine-level similarities in the training and testing sets.

Here, we show that analyzing the similarities between training and testing sets may provide additional insights. We observe that for standard datasets like CIFAR-10 and CIFAR-100, a portion of testing sets have nearly identical samples in the training sets. We also observe that mistakes of some classification models are testing images that are either not similar to any image in the training set, or they are similar to training images with different label.

On the other hand, for the state of the art models that achieve nearly perfect accuracies (e.g., 99.4\% on CIFAR-10), we observe that their few testing mistakes does not seem to be explainable by similarity of images. In fact we see that some of those testing mistakes have nearly identical training samples. We know that image classifiers are highly over-parameterized and there are infinite minimizers for training loss. The art of achieving high accuracy is in fact finding the minimizer of training loss that achieves good accuracy on the testing set. But, can we choose the best minimizer of training loss without looking at the testing set? If one did not have access to testing set of CIFAR-10 dataset, could they pick the model that achieves nearly perfect accuracy on testing set by just learning the training set?

This type of analysis makes it possible to gain insights about the generalization of models for individual images in terms of their similarities with training sets. It also makes it possible to have some measure of confidence about the accuracy of classifications for unlabeled images. For example, if a testing image is nearly identical to some training images of one class, and not similar to training images of any other class, we can be more confident in the accuracy of model's classification. But, if a testing image is dissimilar to all the training set, or it is similar to several images from different classes, we can be less confident in the accuracy of classification for that image.

Moreover, in active learning, an analysis of training set may guide us to acquire images that are less abundant in training set.

\section{Numerical experiments}  \label{sec:results}

Here, we investigate the similarities in three datasets: CIFAR-10, CIFAR-100, and one class of Google Landmarks dataset. The code implementing our methods will be available online.


\subsection{CIFAR-10 dataset}
We use the 2D Daubechies wavelets to decompose all images in this dataset. The matrix of wavelet coefficients has 50,000 rows and 3,072 columns and its condition number is 21,618. We use all the wavelet coefficients, because the condition number of matrix is not very large.

\begin{figure}[h]
\begin{center}
   \fbox{\includegraphics[width=0.2\linewidth]{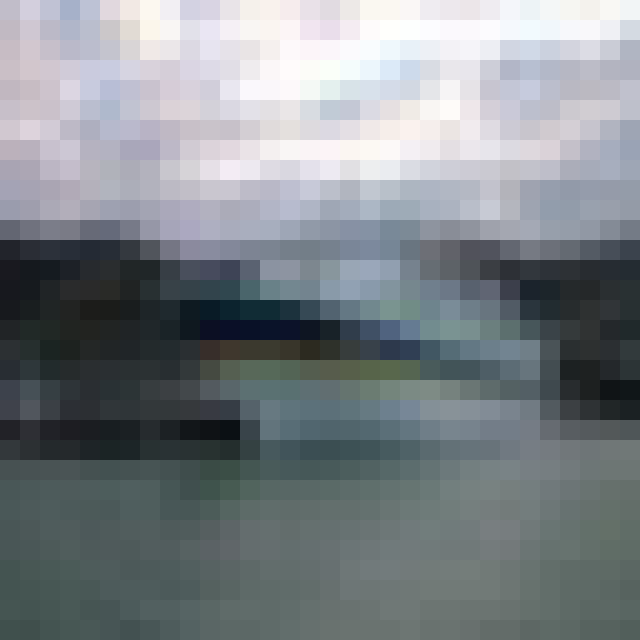}
   \includegraphics[width=0.2\linewidth]{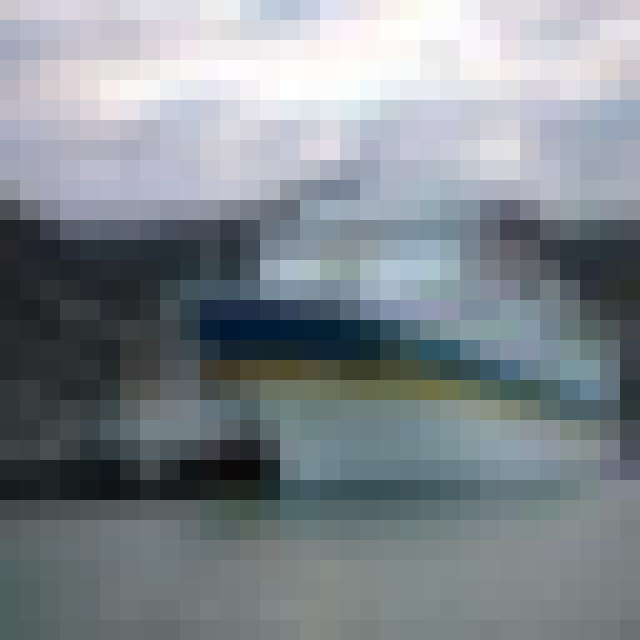}}
   \fbox{\includegraphics[width=0.2\linewidth]{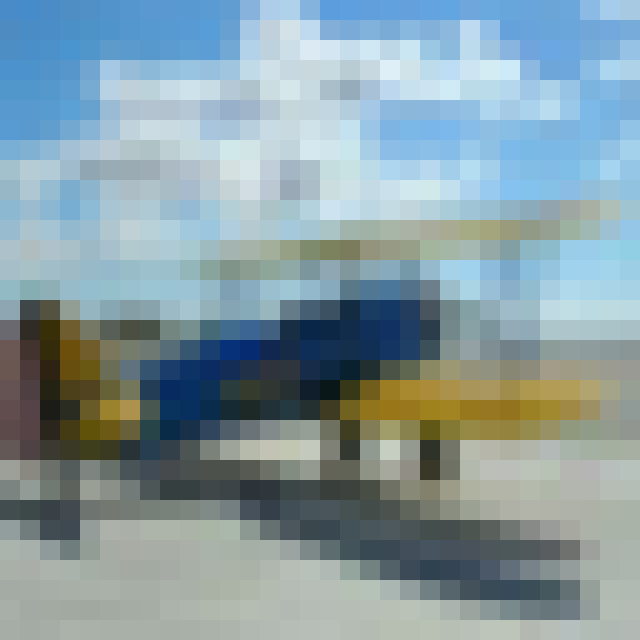}
   \includegraphics[width=0.2\linewidth]{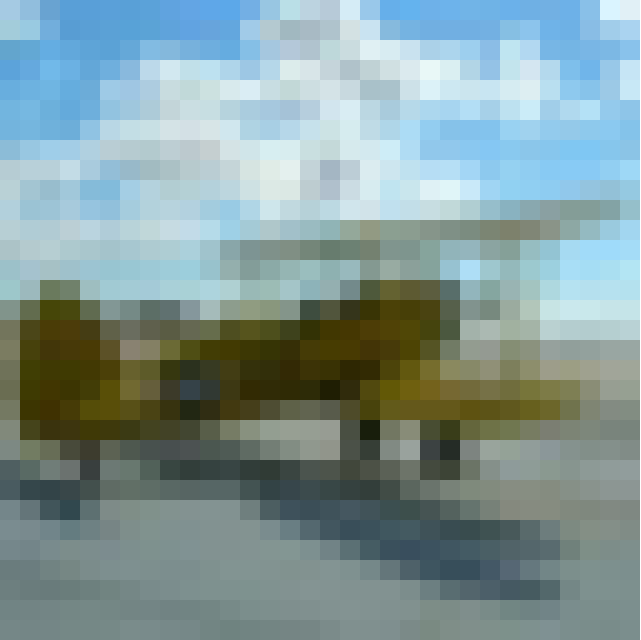}}\\
   \fbox{\includegraphics[width=0.2\linewidth]{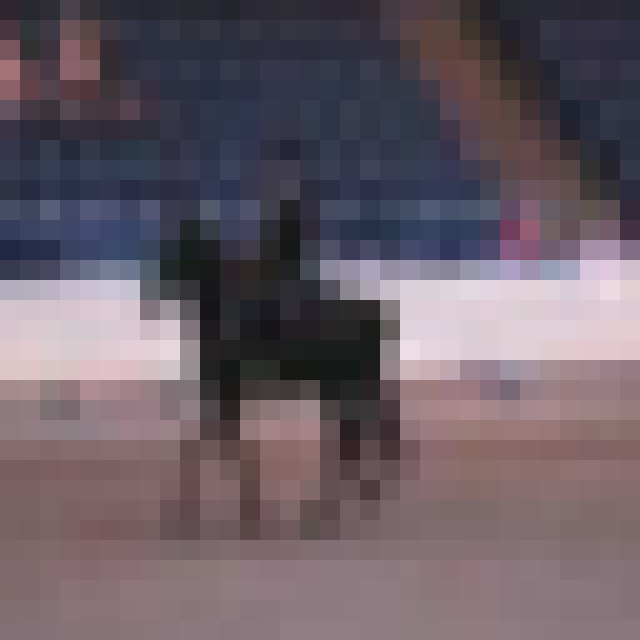}
   \includegraphics[width=0.2\linewidth]{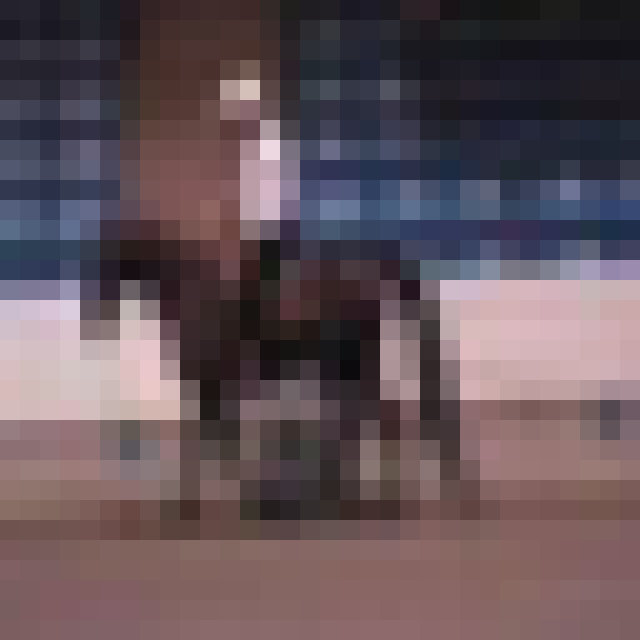}
   \includegraphics[width=0.2\linewidth]{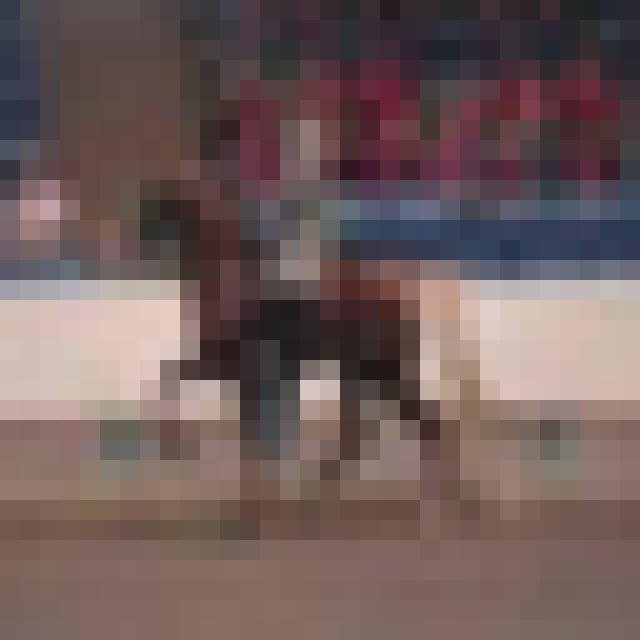}}\\
   \fbox{\includegraphics[width=0.2\linewidth]{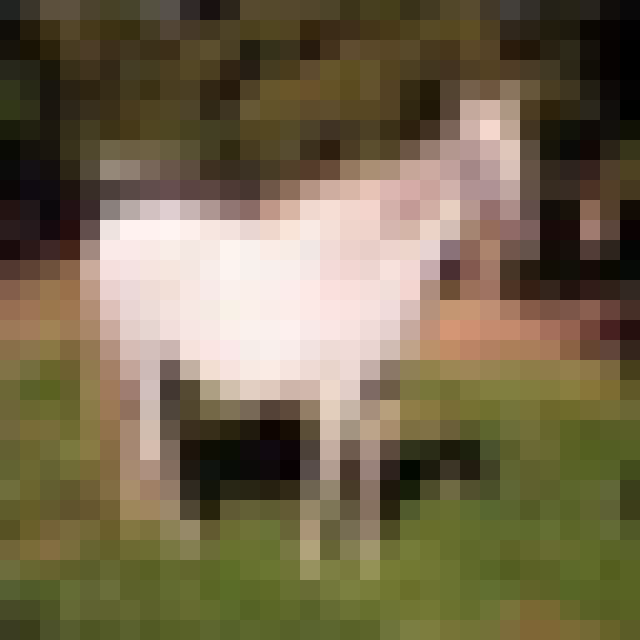}
   \includegraphics[width=0.2\linewidth]{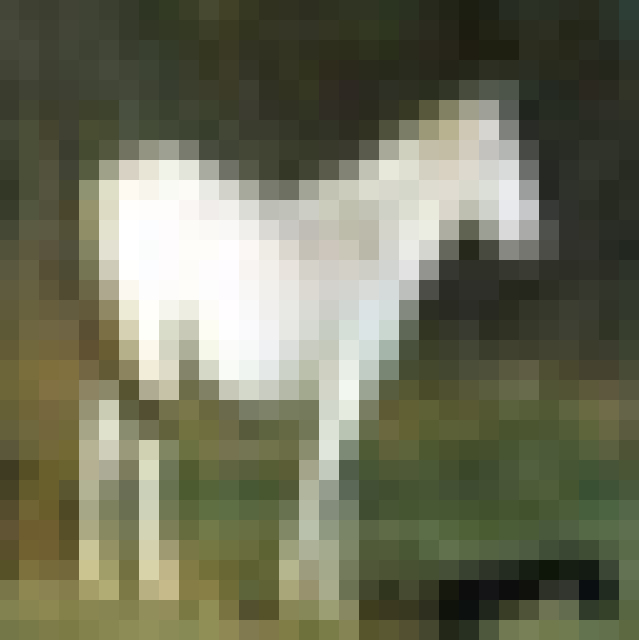}
   \includegraphics[width=0.2\linewidth]{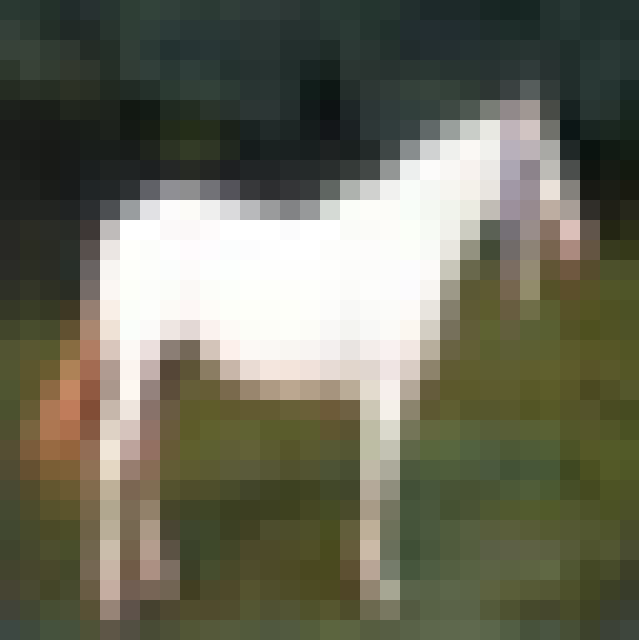}}
\end{center}
   \caption{Example of similar images of same class in CIFAR-10 training set. Images in each box have formed one of the clusters. We see that a standard ResNet model does not have any decision boundaries between images of each group, while it has decision boundaries between dissimilar images of same class.}
\label{fig_cifar10_redund2}
\end{figure}

Using Algorithm~\ref{alg_small}, we cluster the images. We later explain a method for choosing $n_c$ based on an eigen value analysis. But, here we choose $n_c=47,000$ based on the percentage of redundancies reported by \cite{birodkar2019semantic}. Figure~\ref{fig_cifar10_redund2} shows images in some of the clusters with uniform label. Interestingly, the similar images we obtain are the same as the similar images reported by \cite[Appendix]{birodkar2019semantic}. They showed that training a model without the redundant images does not adversely affect the accuracy of models on the testing set. Since our results corroborate their results, there is no need to repeat their experiments on the testing accuracy. One training image per each of these clusters can suffice for training.

The method used by \cite{birodkar2019semantic}, however, requires training a ResNet model on the entire dataset, a process which can take a few GPU hours for the CIFAR-10 dataset. Our contribution here is our fast computational method that identifies similar images very fast. Our other contribution is to report that wavelets identify the similarity of images, in the same way that a pre-trained ResNet-50 does.


Our fast algorithm makes the analysis of contents of image classification models practical, and therefore opens the door to provide additional insights about large datasets, beyond identifying redundancies.

\begin{figure}[h]
\begin{center}
\fbox{\includegraphics[width=0.2\linewidth]{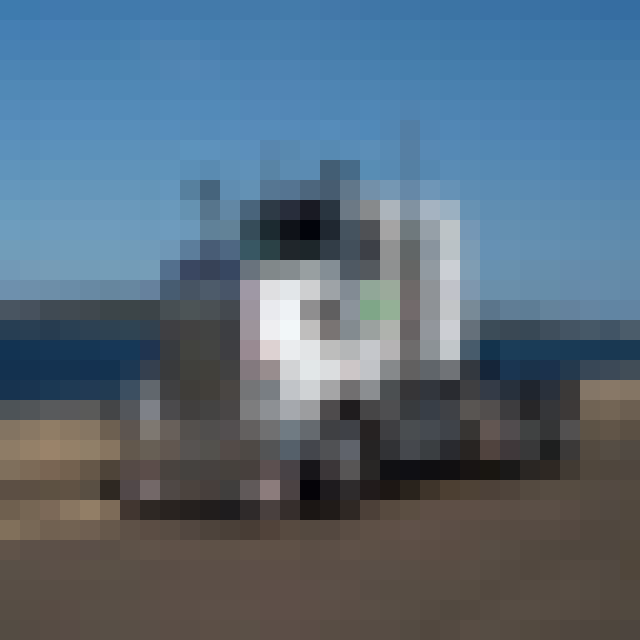}
   \includegraphics[width=0.2\linewidth]{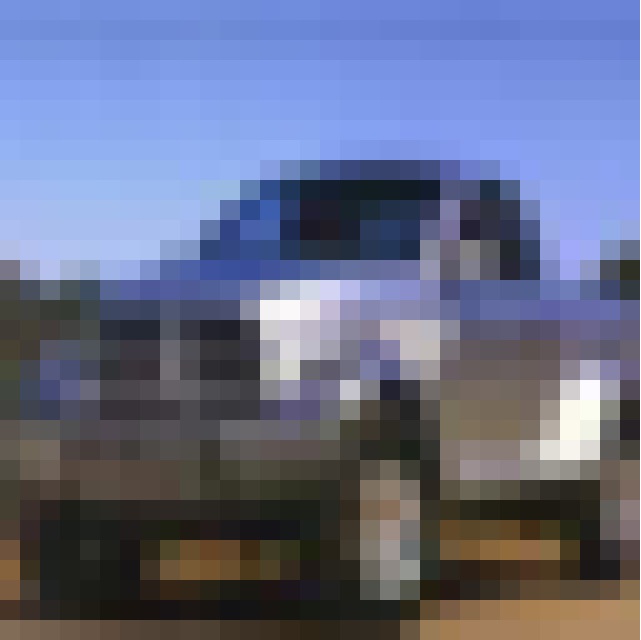}}
\fbox{\includegraphics[width=0.2\linewidth]{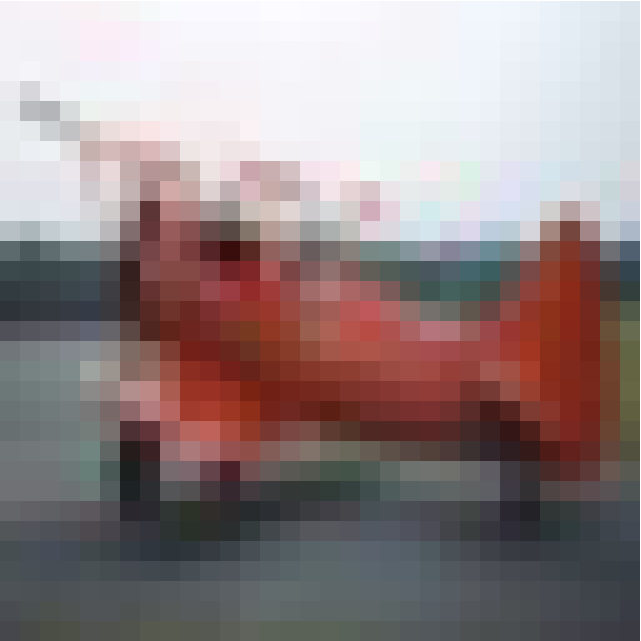}
   \includegraphics[width=0.2\linewidth]{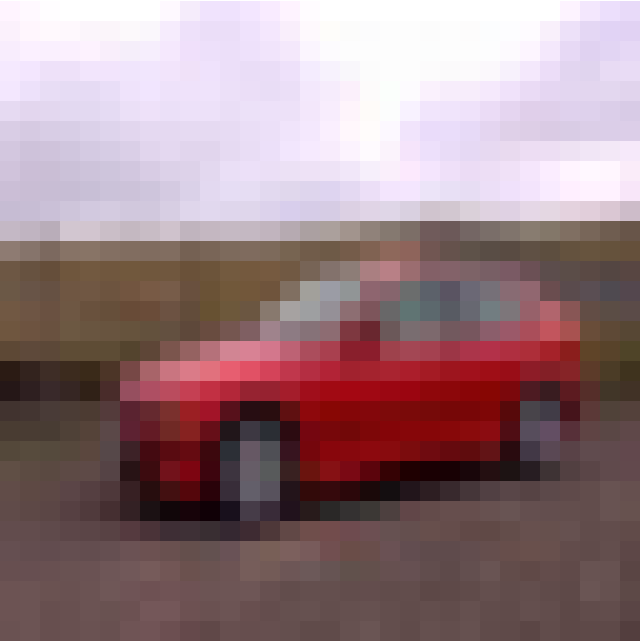}}\\
\fbox{\includegraphics[width=0.2\linewidth]{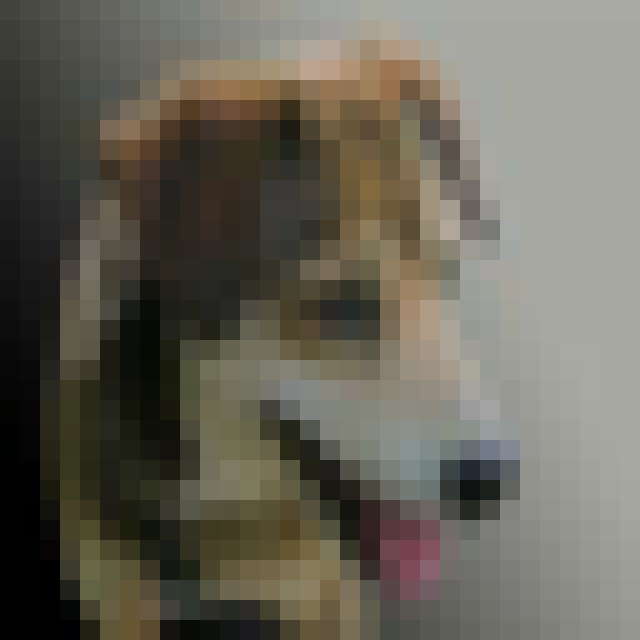}
   \includegraphics[width=0.2\linewidth]{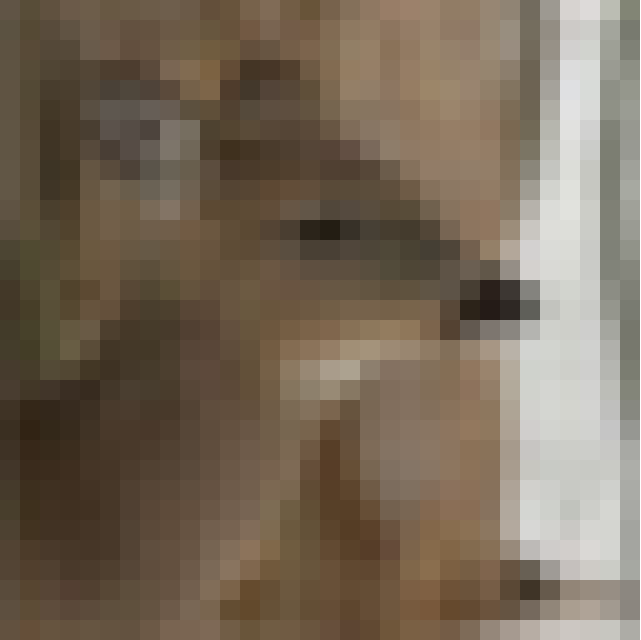}}
\fbox{\includegraphics[width=0.2\linewidth]{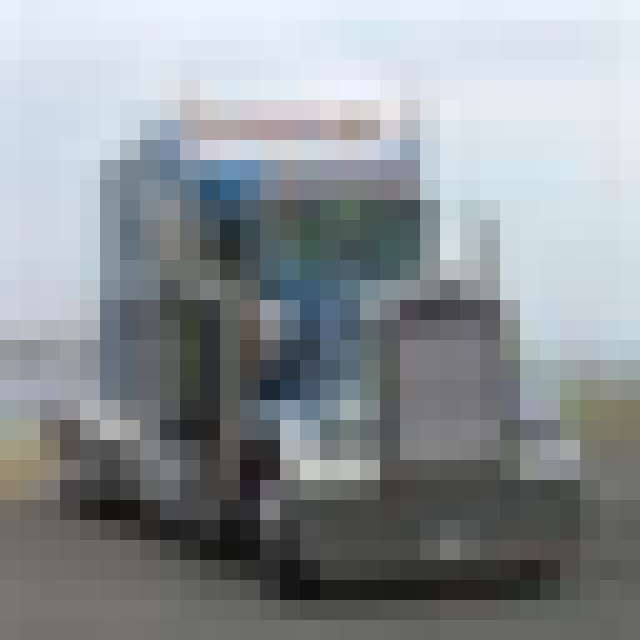}
   \includegraphics[width=0.2\linewidth]{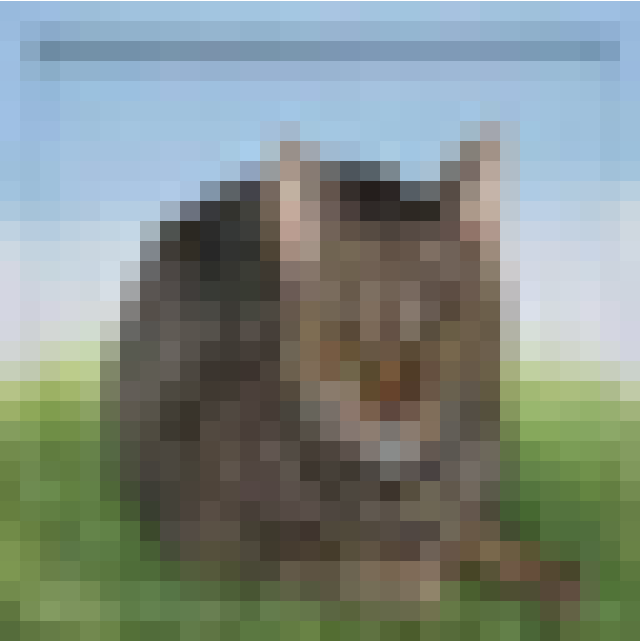}}
\end{center}
   \caption{Some of the similar images with different labels in CIFAR-10 training set. Each box shows one cluster. We consider these images influential in learning.
}
\label{fig_cifar10_inf}
\end{figure}

Figure~\ref{fig_cifar10_inf} shows some of the same cluster images that have different labels. Clearly, it is desirable for a model to learn these images and be able to distinguish them from each other.

\subsection{CIFAR-100 dataset}

\begin{figure}[h]
\begin{center}
   \fbox{\includegraphics[width=0.2\linewidth]{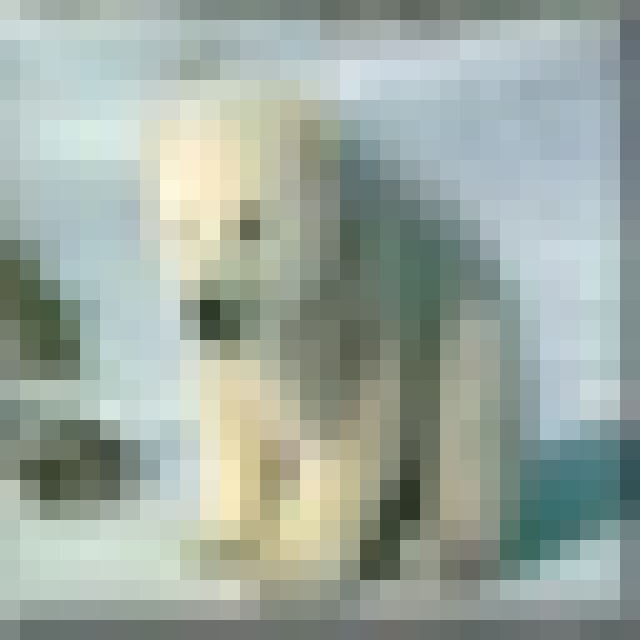}
   \includegraphics[width=0.2\linewidth]{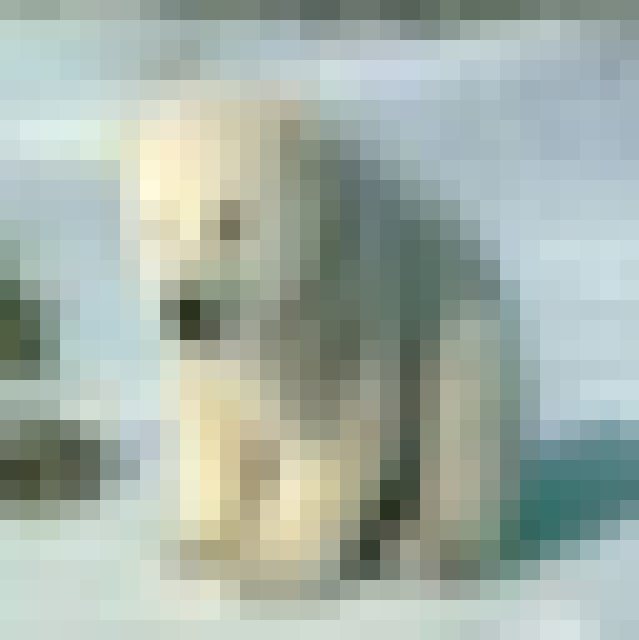}
   \includegraphics[width=0.2\linewidth]{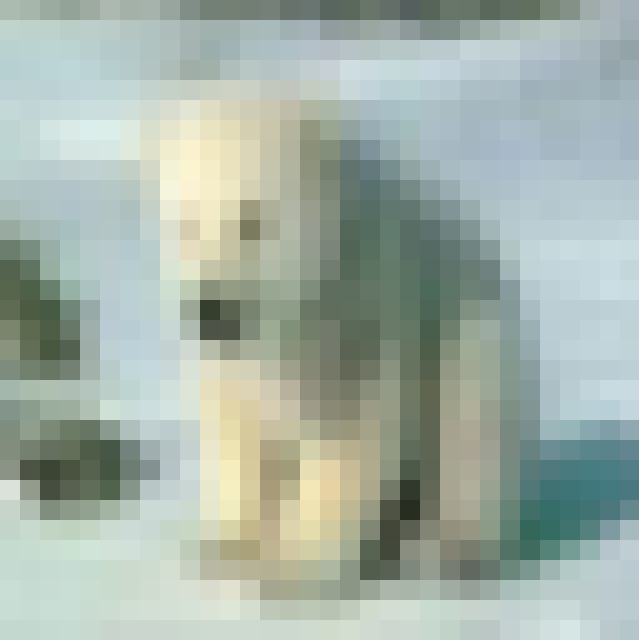}}
\end{center}
   \caption{Example of redundant images in CIFAR-100 training set.}
\label{fig_cifar100_redund}
\end{figure}

We identify redundancies in the CIFAR-100 training set, too, as briefly shown in Figure~\ref{fig_cifar100_redund}. 

Similar images with different labels are abundant in this dataset and might be even hard to distinguish for a human. For example, the image pair in the left box in Figure~\ref{fig_cifar100_inf} represent a maple tree and an oak tree, and the image pair in the right box represent a whale and a seal.



\subsubsection{Class of aquarium fish (training set)}
To gain more insight and to compare our algorithms, here, we consider only the second class of this dataset with 5,000 images. 

\begin{figure}[h]
\begin{center}
   \fbox{\includegraphics[width=0.2\linewidth]{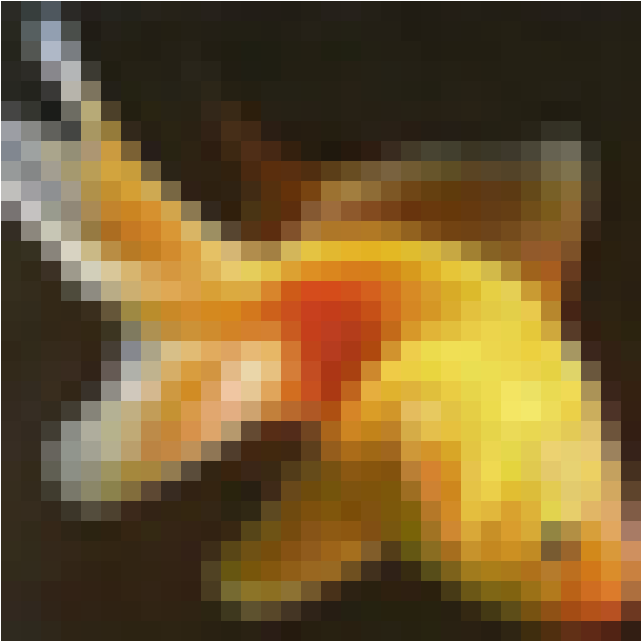}
   \includegraphics[width=0.2\linewidth]{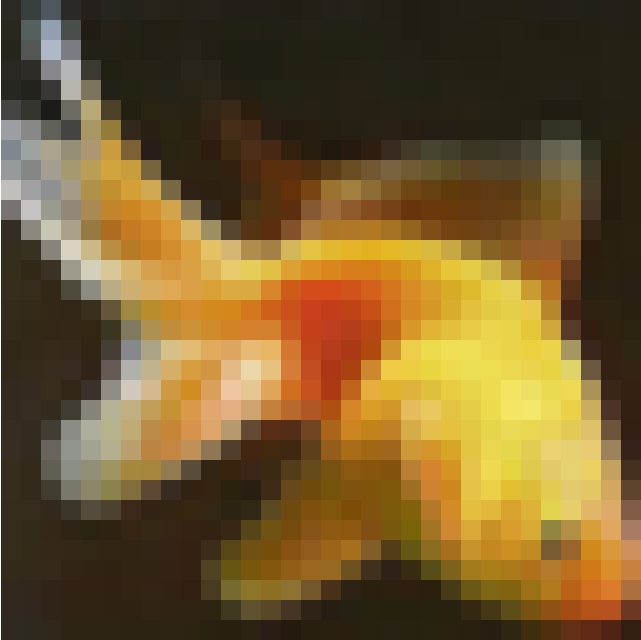}}
   \fbox{\includegraphics[width=0.2\linewidth]{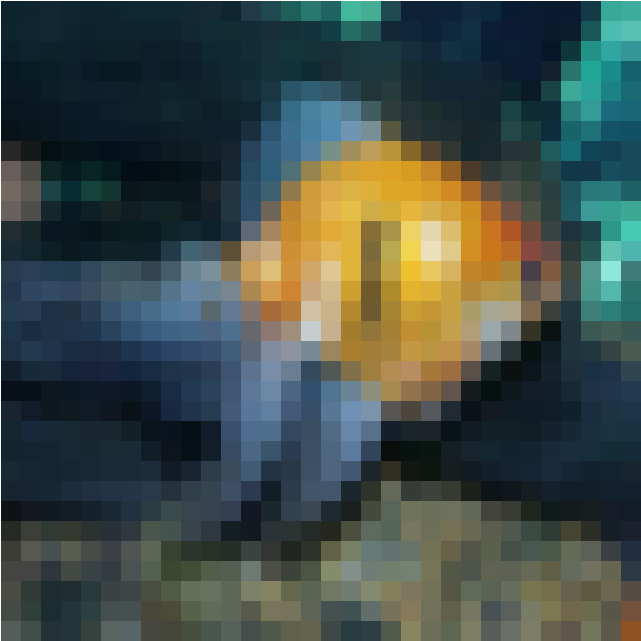}
   \includegraphics[width=0.2\linewidth]{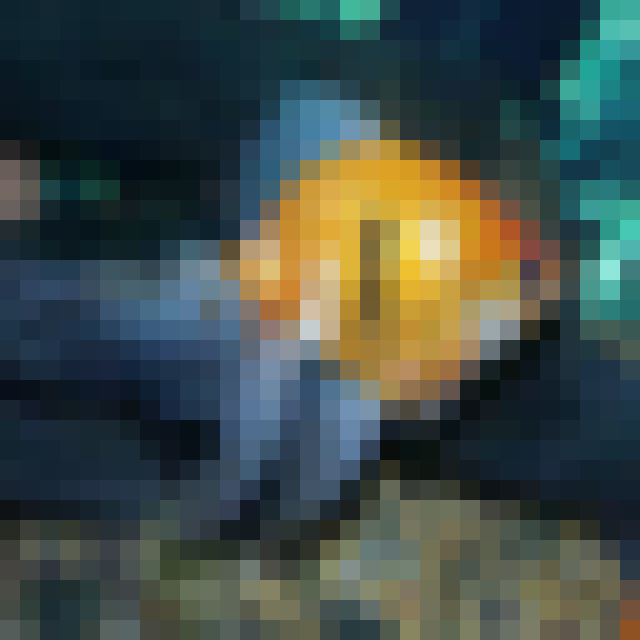}}
\end{center}
   \caption{Example of redundant images in the aquarium fish class of CIFAR-100.}
\label{fig_cifar100_redund_fish}
\end{figure}

Starting by Algorithm~\ref{alg_small}, the matrix of wavelet coefficients for this class is $5,000\times 3,072$, with condition number $4\times 10^{18}$. Numerical rank of this matrix is 495, using rank tolerance of $\tau = 10^{-5}$. We identify the 495 wavelet coefficients using rank-revealing QR factorization and use them for clustering with $n_c = 470$. The entire computation takes about 5 seconds on a machine with a 2.30GHz CPU and 115GB of RAM. We obtain redundant images as shown in Figure~\ref{fig_cifar100_redund_fish}.

\begin{figure}[h]
\begin{center}
   \fbox{\includegraphics[width=0.2\linewidth]{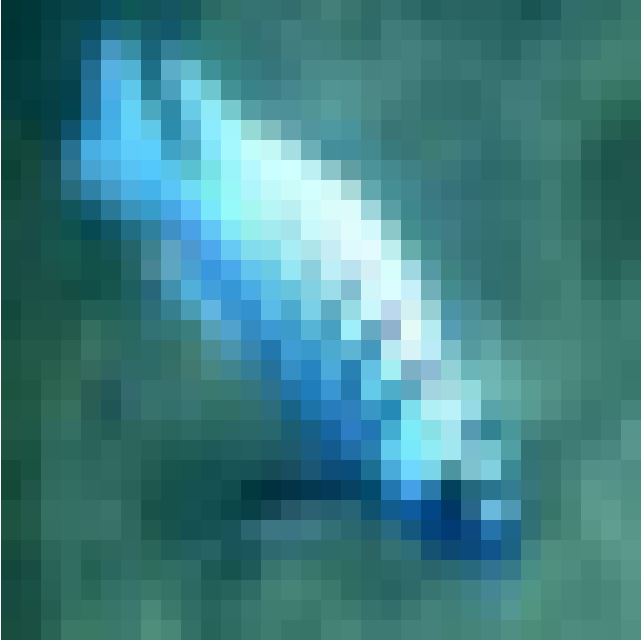}
   \includegraphics[width=0.2\linewidth]{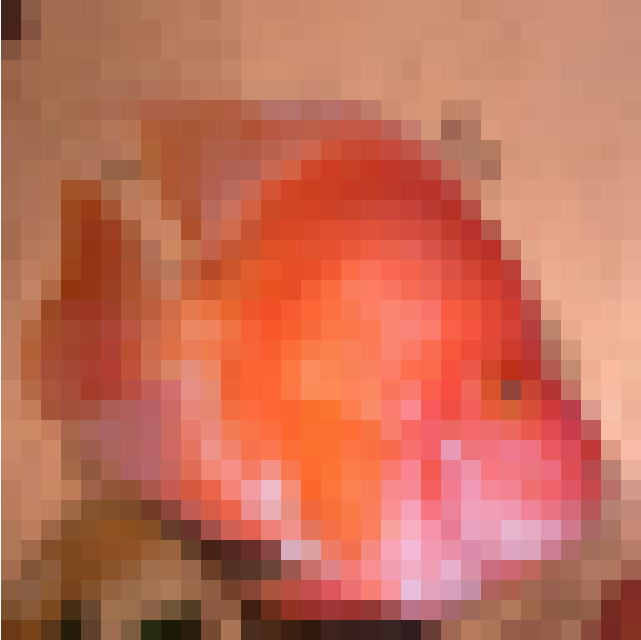}}
   \fbox{\includegraphics[width=0.2\linewidth]{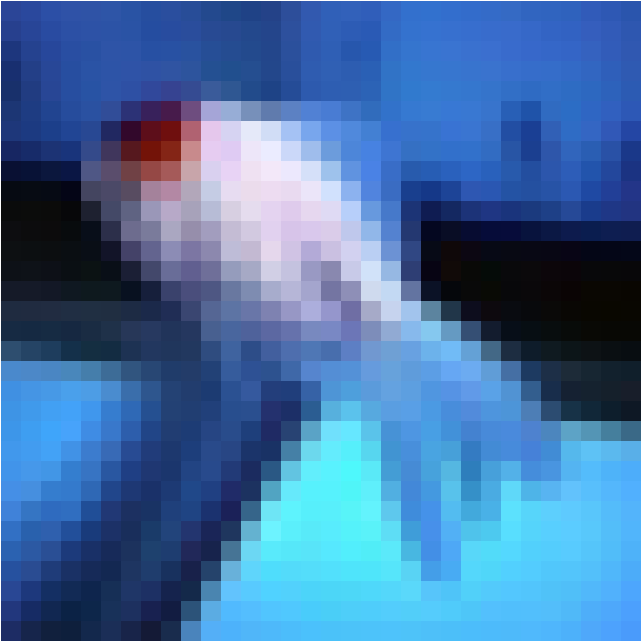}
   \includegraphics[width=0.2\linewidth]{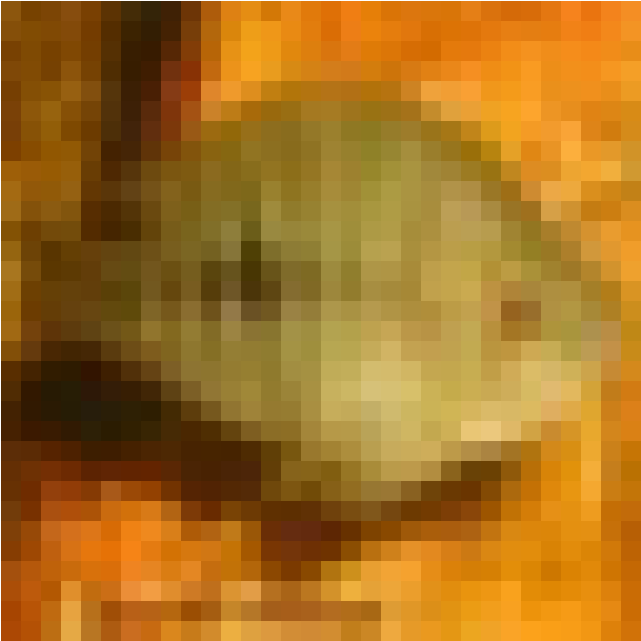}}
\end{center}
   \caption{Two most dissimilar image pairs in the aquarium fish class of CIFAR-100, based on the SSIM measure. SSIM is -0.5420 for images in the left box and -0.5025 for the right box.}
\label{fig_cifar100_dissim_fish}
\end{figure}

Let's see how Algorithm~\ref{alg_spectral} performs and what additional information it can provide. 
According to the SSIM measure, there are 5 pairs of identical images, all of which are also picked by Algorithm~\ref{alg_small} (shown in Figure~\ref{fig_cifar100_redund_fish}). In contrast, Figure~\ref{fig_cifar100_dissim_fish} shows two of the most dissimilar image pairs based on the SSIM measure.

\begin{figure}[h]
\begin{center}
   \includegraphics[width=0.99\linewidth]{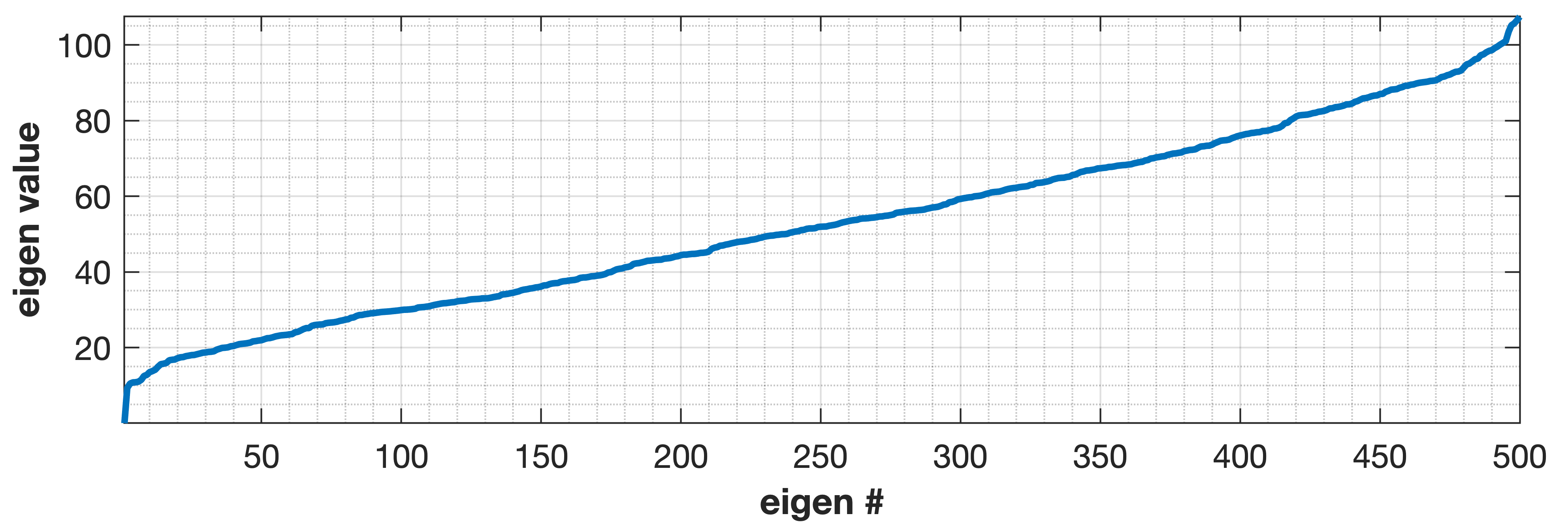}
\end{center}
   \caption{Distribution of eigenvalues of the graph Laplacian for all images in the ``aquarium fish" class of CIFAR-100.}
\label{fig_cifar100_spectrum}
\end{figure}

The mean value of the similarity matrix, $\mathcal{S}$, is 0.088 and its standard deviation is 0.131. Figure~\ref{fig_cifar100_spectrum} shows the distribution of eigenvalues of its graph Laplacian, implying that there are not any large clusters in the data. We choose the number of clusters based on the eigen-gaps of the graph Laplacian. Using the eigen-gap threshhold as 0.4 leads to $n_c = 454$.

\begin{figure}[h]
\begin{center}
   \fbox{\includegraphics[width=0.2\linewidth]{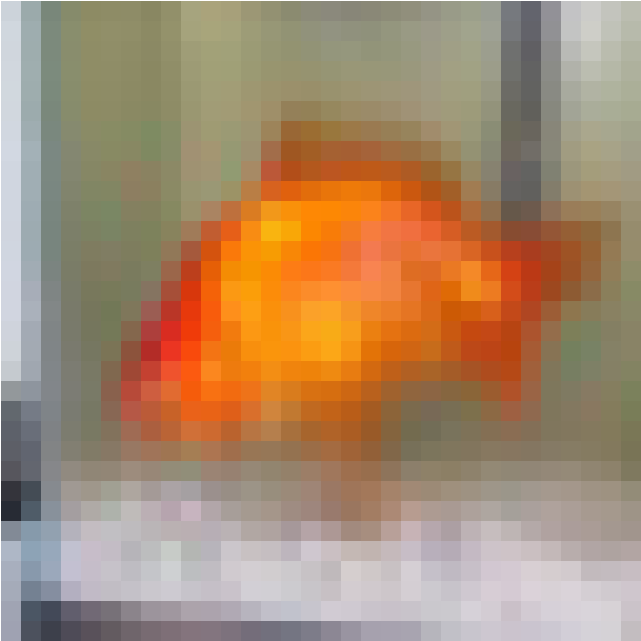}
   \includegraphics[width=0.2\linewidth]{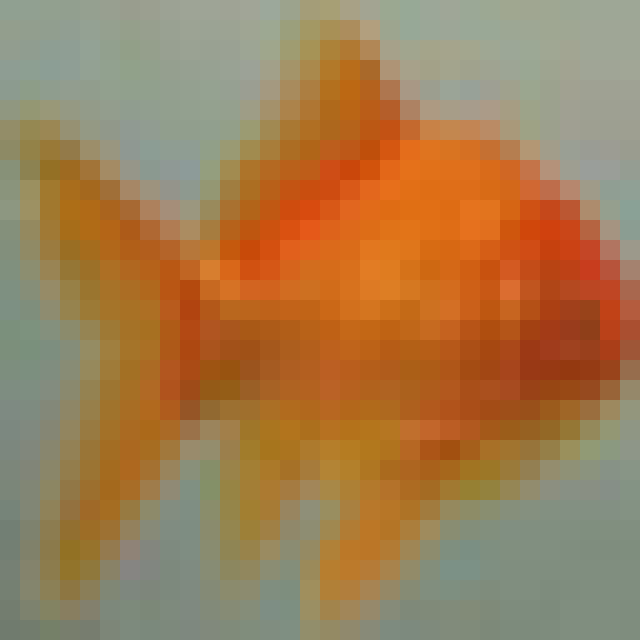}}
   \fbox{\includegraphics[width=0.2\linewidth]{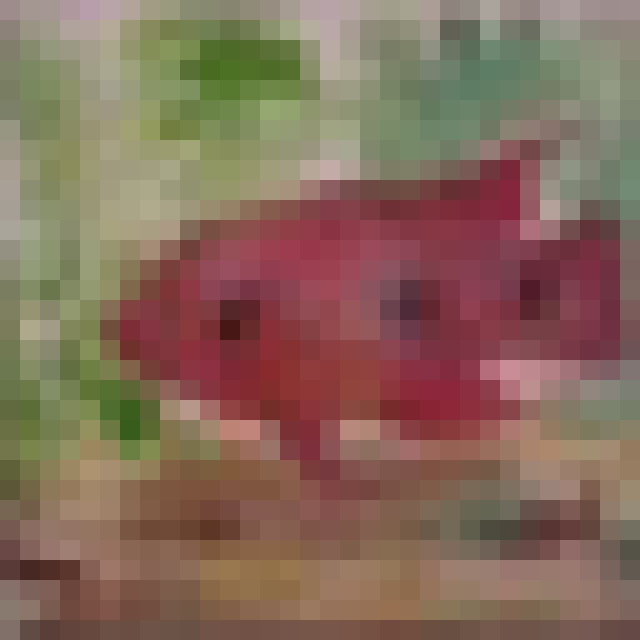}
   \includegraphics[width=0.2\linewidth]{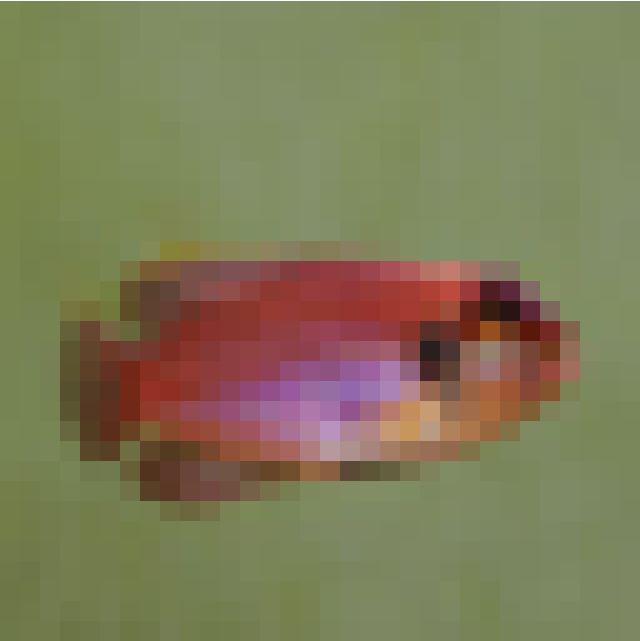}}
\end{center}
   \caption{Examples of similar training images in CIFAR-100.}
\label{fig_cifar100_sim_spect}
\end{figure}

Spectral clustering then yields clusters with all the identical pairs mentioned above, with some additional images that are fairly similar, as two pairs are shown in Figure~\ref{fig_cifar100_sim_spect} because we chose a smaller $n_c$.

In summary, the results of our two algorithms corroborate each other. Algorithm~\ref{alg_spectral} is more expensive for this example as expected, however, it provides more detailed insights about the images and guides us to choose a wise value for the number of clusters. 

\subsection{Generalization of models}
\subsubsection{Class of aquarium fish of CIFAR-100 (training and testing sets)}

\begin{figure}[h]
\begin{center}
   \includegraphics[width=0.95\linewidth]{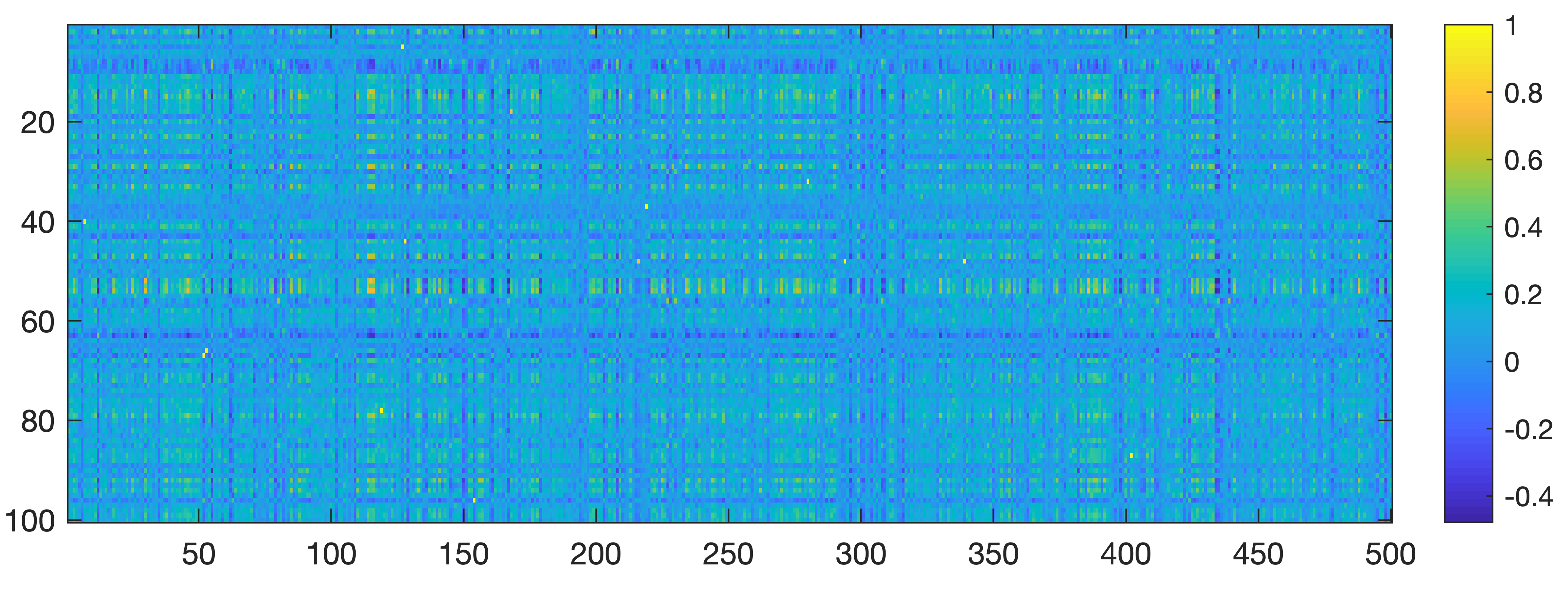}
\end{center}
   \caption{The similarity matrix between 100 testing images and 500 training images of the aquarium fish class in CIFAR-100.}
\label{fig_cifar100_fish_matrix_tt}
\end{figure}

\begin{figure}[h]
\begin{center}
   \includegraphics[width=0.2\linewidth]{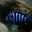}
   \includegraphics[width=0.2\linewidth]{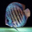}
   \includegraphics[width=0.2\linewidth]{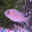}
\end{center}
   \caption{Testing images most dissimilar from the training set for the aquarium fish class. These happen to be common mistakes of some models.}
\label{fig_cifar100_fish_dissim}
\end{figure}

We compare all 100 testing images of this class to all 500 training images of this class. The similarity matrix is shown in Figure~\ref{fig_cifar100_fish_matrix_tt}. This analysis shows that 11\% of testing images have a nearly identical image in training set. Figure~\ref{fig_cifar100_fish_dissim} shows three testing images that have the least similarity to all images in the training set which are among the mistakes of some classification models on CIFAR-100.

\subsubsection{Classes of cat and dog of CIFAR-10 (training and testing sets)}

We consider the model developed by \citet{kolesnikov2019large} which has reported the best accuracy on CIFAR-10. This model only makes 65 misclassifications out of the entire 10,000 testing images of CIFAR-10 dataset. 19 of those mistakes are either misclassifying a dog as a cat, or the reverse. So, we consider those two classes and analyze the similarities between their training and testing images.

In this case, we decompose images using the Daubechies 2 wavelets, measure their distance in Euclidean space, and then convert the distance to a similarity measure using Gaussian kernels. 

\begin{figure}[h]
\begin{center}
   \includegraphics[width=0.2\linewidth]{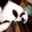}
   \includegraphics[width=0.2\linewidth]{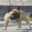}
   \includegraphics[width=0.2\linewidth]{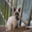}
   \includegraphics[width=0.2\linewidth]{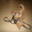}
   \includegraphics[width=0.2\linewidth]{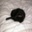}
   \includegraphics[width=0.2\linewidth]{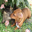}
   \includegraphics[width=0.2\linewidth]{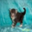}
\end{center}
   \caption{Testing images of cat in CIFAR-10 that are misclassified by the state of art model.}
\label{fig_cifar10_cat_mistakes}
\end{figure}

\begin{figure}[h]
\begin{center}
   \fbox{\includegraphics[width=0.2\linewidth]{cifar_test_mist_1502.png}}
   \fbox{\includegraphics[width=0.2\linewidth]{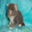}
   \includegraphics[width=0.2\linewidth]{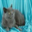}
   \includegraphics[width=0.2\linewidth]{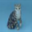}}
\end{center}
   \caption{(left) A testing image misclassified by the state of art model on CIFAR-10. (right) Three training images most similar to the image on left, all labeled as cat.}
\label{fig_cifar10_cat_simto1502}
\end{figure}

Testing images of Cat that are misclassified as Dog by \cite{kolesnikov2019large} are shown in Figure~\ref{fig_cifar10_cat_mistakes}. Consider the image at the bottom right in this Figure. The three most similar training images to it (from both Cat and Dog classes) are shown in Figure~\ref{fig_cifar10_cat_simto1502}, all of which have cat label and are considerably similar to the misclassified testing image. So, absence of similar training data with same label, nor presence of similar training data with opposite label may not be the cause of misclassification. 

\begin{figure}[h]
\begin{center}
   \includegraphics[width=0.2\linewidth]{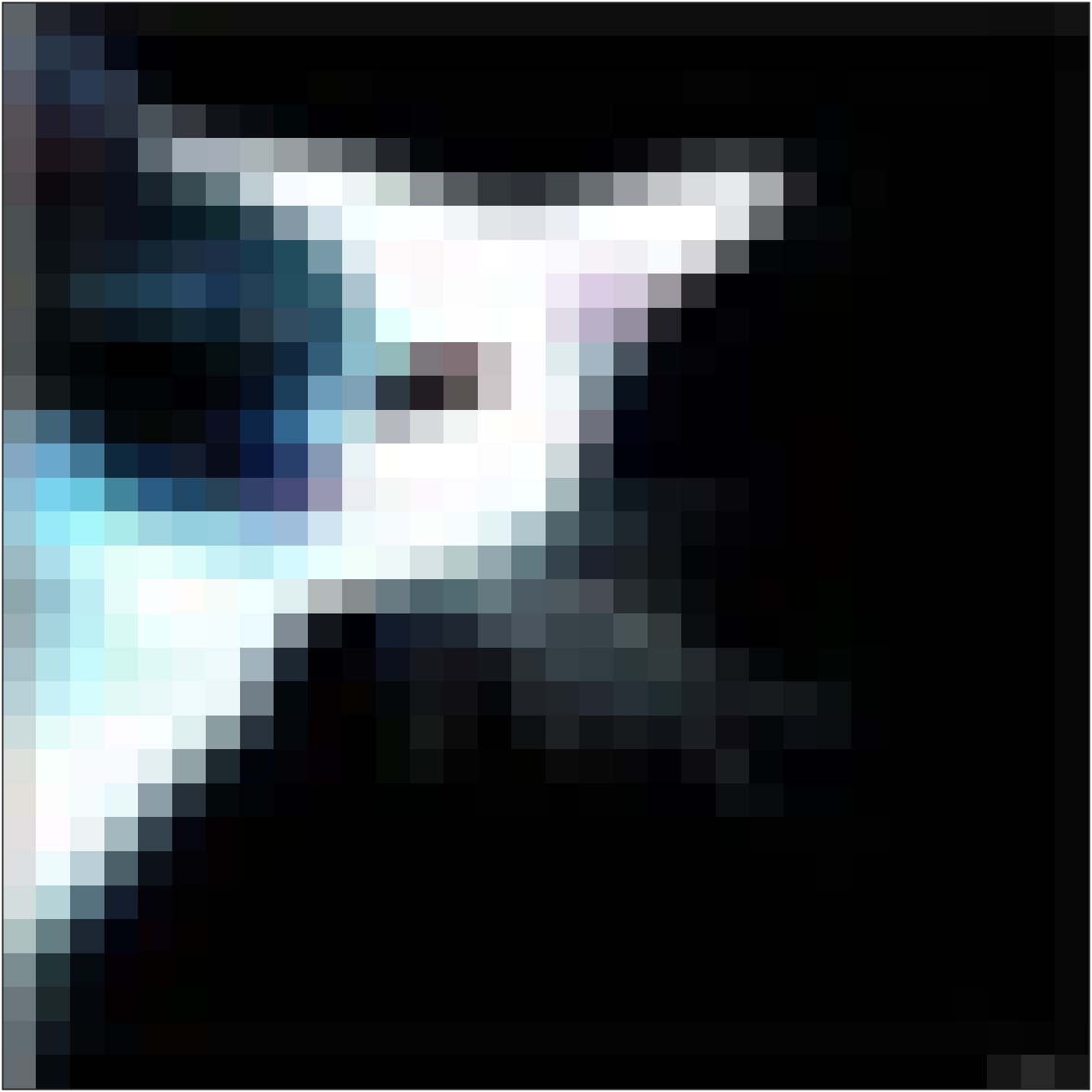}
   \includegraphics[width=0.2\linewidth]{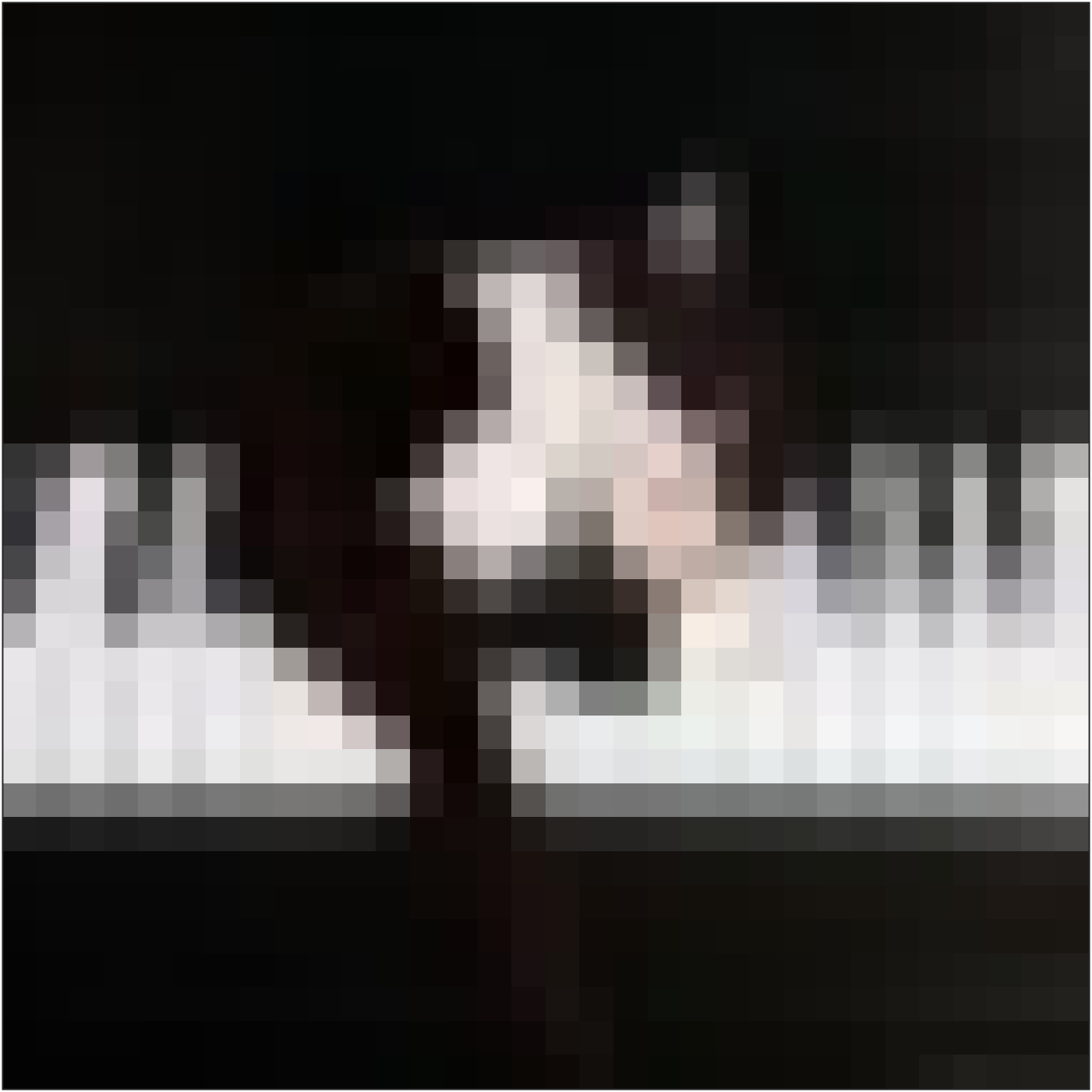}
   \includegraphics[width=0.2\linewidth]{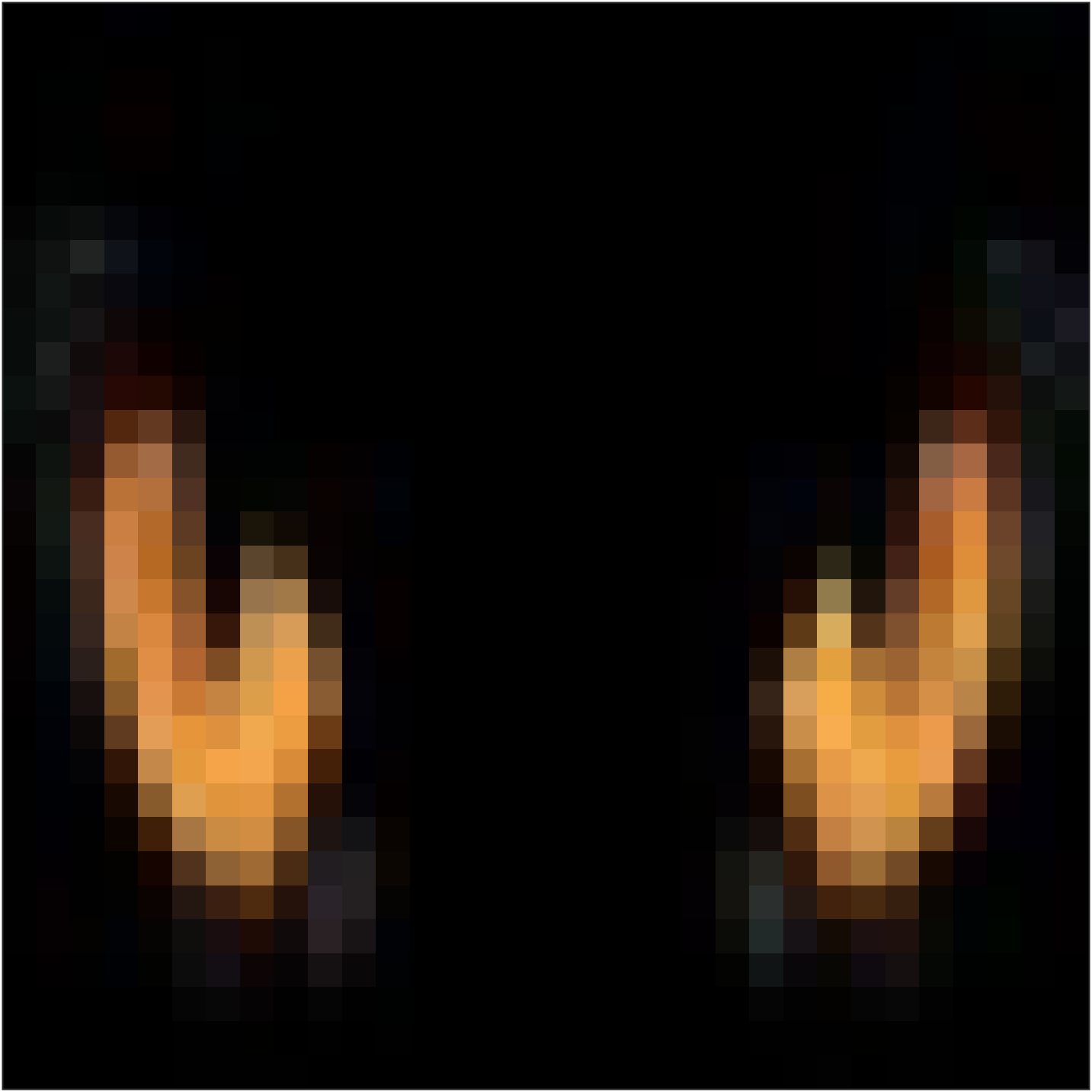}
\end{center}
   \caption{Testing images most dissimilar from the training set for the cat class of CIFAR-10.}
\label{fig_cifar10_cat_strange}
\end{figure}

On the other hand, there are correctly classified testing images that can be considered isolated from the training set. Figure~\ref{fig_cifar10_cat_strange} shows the three testing images of Cat class that are most dissimilar to the entire training set of cats and dogs. The model developed by \cite{kolesnikov2019large} correctly classifies them as cat.

\begin{figure}[h]
\begin{center}
   \includegraphics[width=0.2\linewidth]{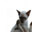}
   \includegraphics[width=0.2\linewidth]{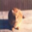}
   \includegraphics[width=0.2\linewidth]{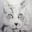}
\end{center}
   \caption{Testing images of cat in CIFAR-10 that are more similar to training images of dogs, compared to training images of cats.}
\label{fig_cifar10_cat_simdog}
\end{figure}

Moreover, there are testing images of cat that are more similar to training images of dogs compared to training images of cats. Figure~\ref{fig_cifar10_cat_simdog} shows three of those, all of which are correctly classified as cat by \cite{kolesnikov2019large}.

It is becoming common in the literature to conduct surveys about the mistakes of the models and ask humans whether they can classify them correctly, in order to justify the mistakes. However to our knowledge, surveys are not conducted about correctly classified testing images. In fact one might wonder why the image in the middle of Figure~\ref{fig_cifar10_cat_strange} and the rightmost image in Figure~\ref{fig_cifar10_cat_simdog} are labeled as cat.

We hope that these observations lead to meaningful questions and answers about the generalization in deep learning.

\subsection{Influence of training data on decision boundaries of a trained model}

We consider the standard ResNet-v2 models \citep{he2016identity}, pre-trained on CIFAR-10 and CIFAR-100 datasets and investigate their decision boundaries in relation to these images. A model's decision boundary between two class is any point that produces equal softmax scores for those, while the softmax score for all other classes are less than those \citep{elsayed2018large,yousefzadeh2019interpreting}.

We aim to find whether the output of the model along the direct path connecting two images hits a decision boundary or not. In other words, we want to find out whether the model has a decision boundary defined between two images. The direct path between two images $\bf{x_1}$ and $\bf{x_2}$ is defined by $(1-\alpha)\bf{x_1}+ \alpha\bf{x_2}$, where $\alpha$ is a scalar between 0 and 1.

As expected, images that are almost identical in Figures~\ref{fig_cifar10_redund},\ref{fig_cifar100_redund}, and \ref{fig_cifar100_redund_fish}, do not have any decision boundary between them. On the other hand, images of the same class that are not similar do have decision boundaries between them, for example, images in Figure~\ref{fig_cifar100_dissim_fish}. This means that the model output along the direct path between such images exits the correct classification and re-enters it, hitting at least two decision boundaries in between. Interestingly, groups of images in Figures~\ref{fig_cifar10_redund2} and \ref{fig_cifar100_sim_spect} that are similar but not identical, do not have any decision boundaries between them. This can be the subject of further study.


\subsection{Google Landmarks dataset v2}

\begin{figure}[h]
\begin{center}
   \includegraphics[width=0.55\linewidth]{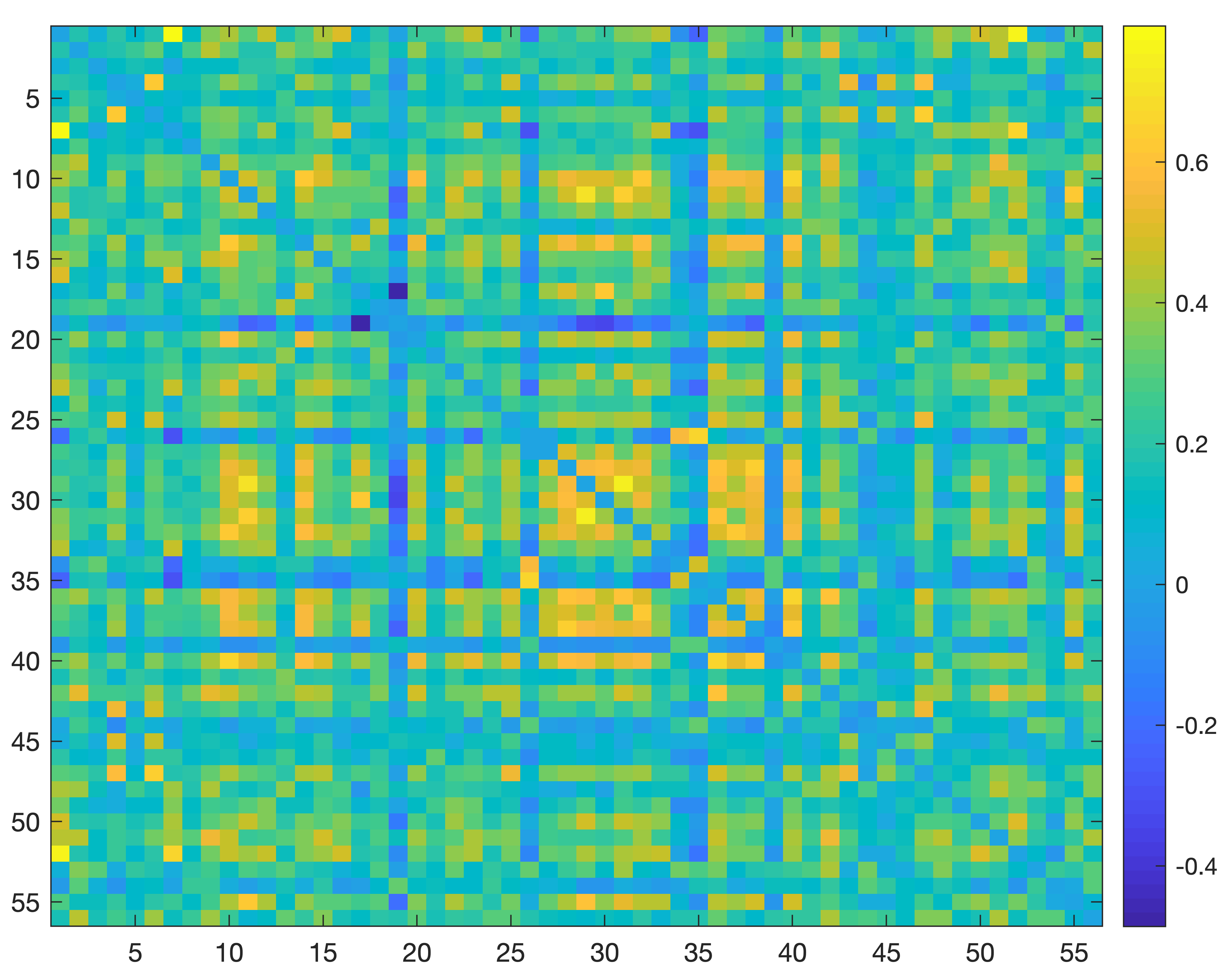}
\end{center}
   \caption{The similarity matrix for the class of Verrazzano-Narrows bridge in the Google Landmarks dataset v2.}
\label{fig_landmark_matrix}
\end{figure}

\begin{figure}[h]
\begin{center}
   \includegraphics[width=0.99\linewidth]{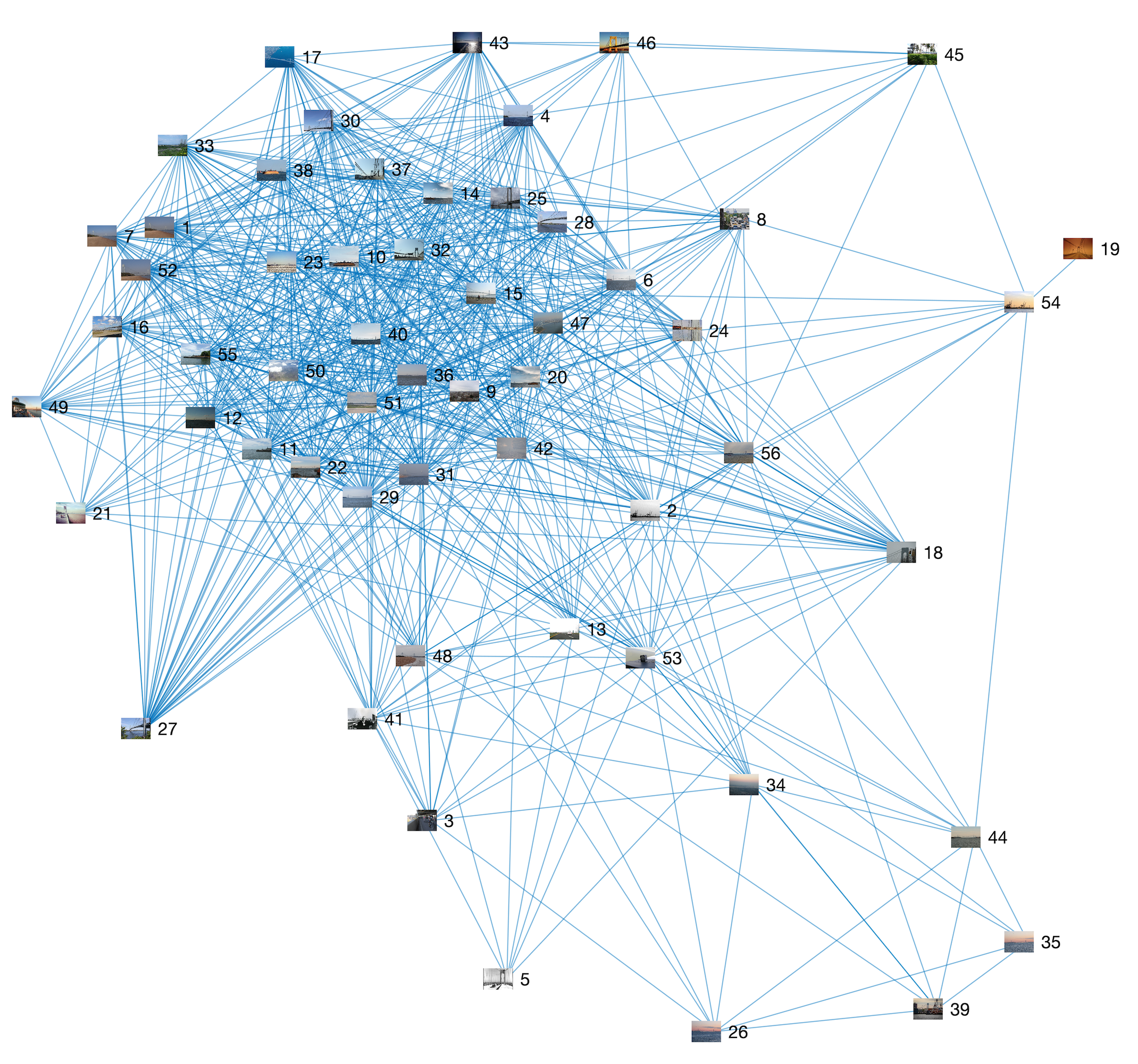}
\end{center}
   \caption{The graphical model derived from the similarity matrix.}
\label{fig_landmark_gm}
\end{figure}

For this dataset, we consider the class of Verrazzano-Narrows bridge. There are 56 images for this class which we standardized as 512 by 662 pixels. Using Algorithm~\ref{alg_spectral}, we analyze all training images in this class. Figure~\ref{fig_landmark_matrix} shows the similarity matrix of images in the class and Figure~\ref{fig_landmark_gm} shows a graphical model derived from the similarity matrix.

\begin{figure}[h]
\begin{center}
   \includegraphics[width=0.32\linewidth]{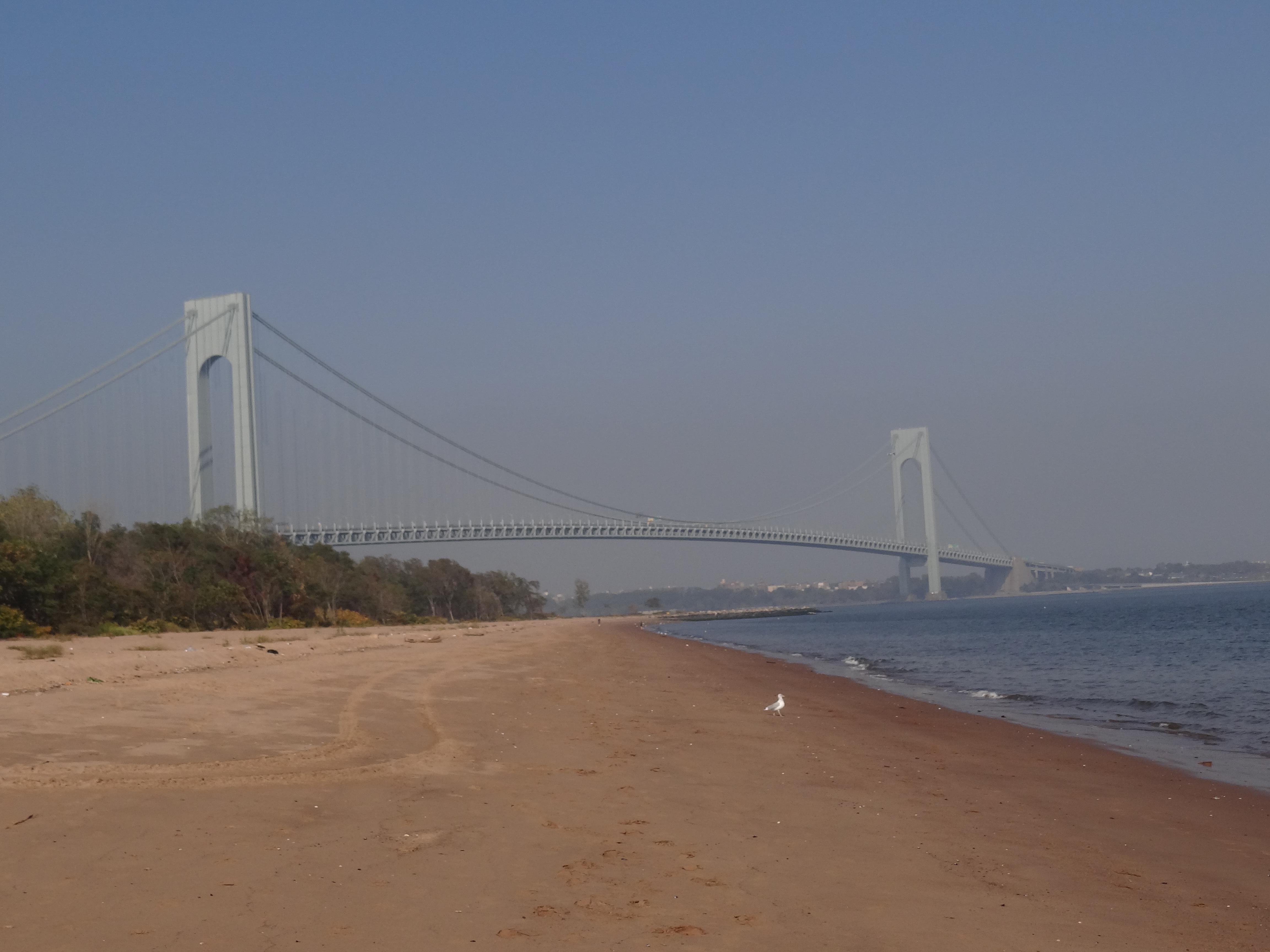}
   \includegraphics[width=0.32\linewidth]{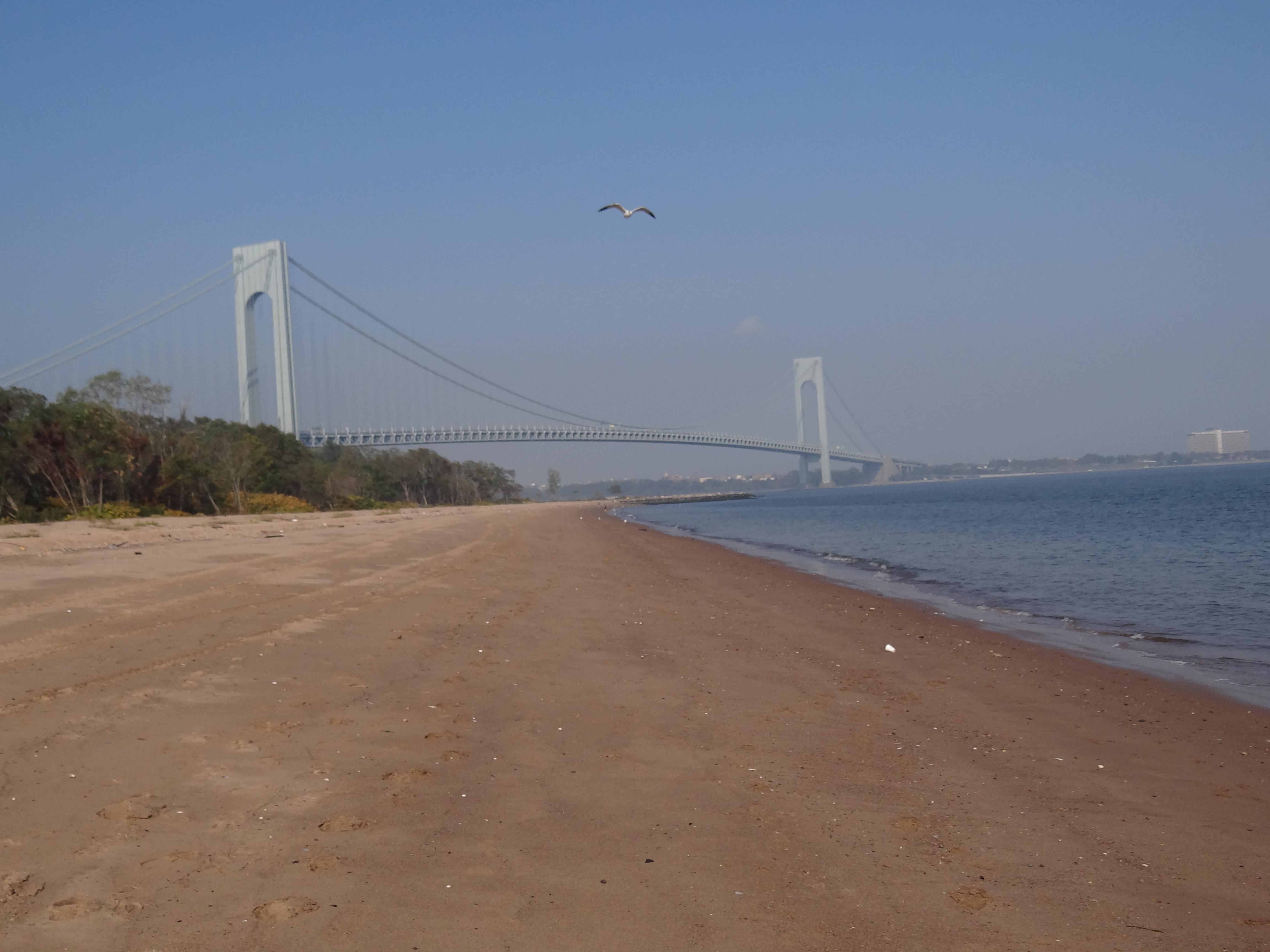}
   \includegraphics[width=0.32\linewidth]{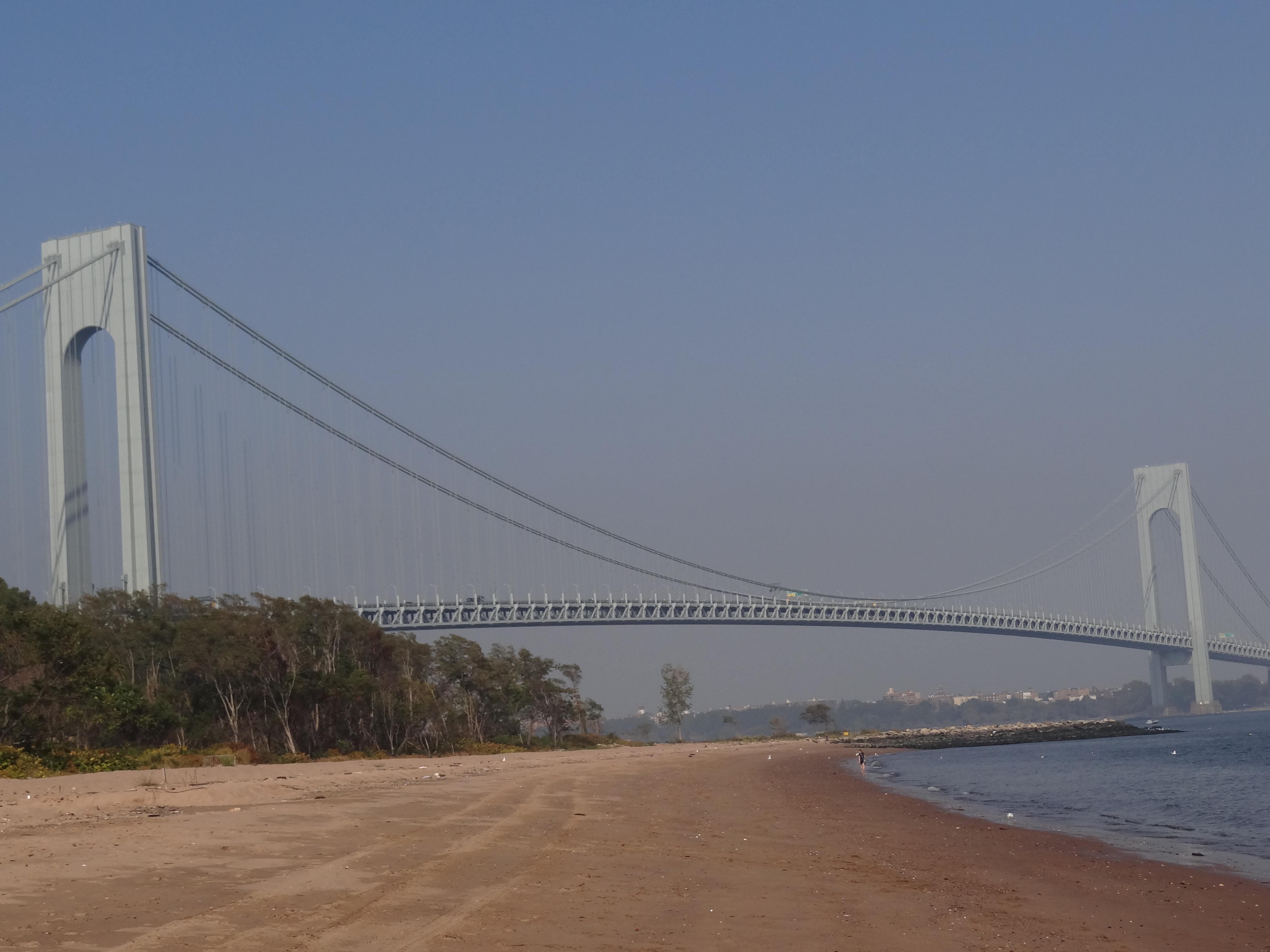}
\end{center}
   \caption{Group of most similar images in the class (images 1,7, and 52 in the graphical model shown in Figure~\ref{fig_landmark_gm}).}
\label{fig_landmark_redund}
\end{figure}

\begin{figure}[H]
\begin{center}
   \includegraphics[width=0.32\linewidth]{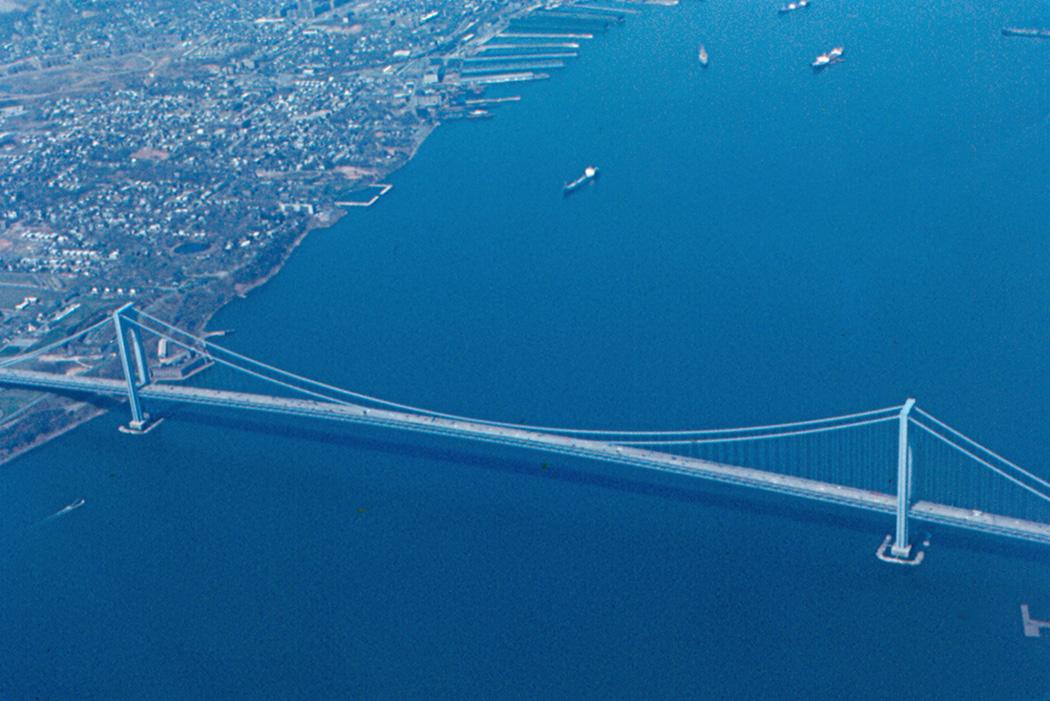}
   \includegraphics[width=0.32\linewidth]{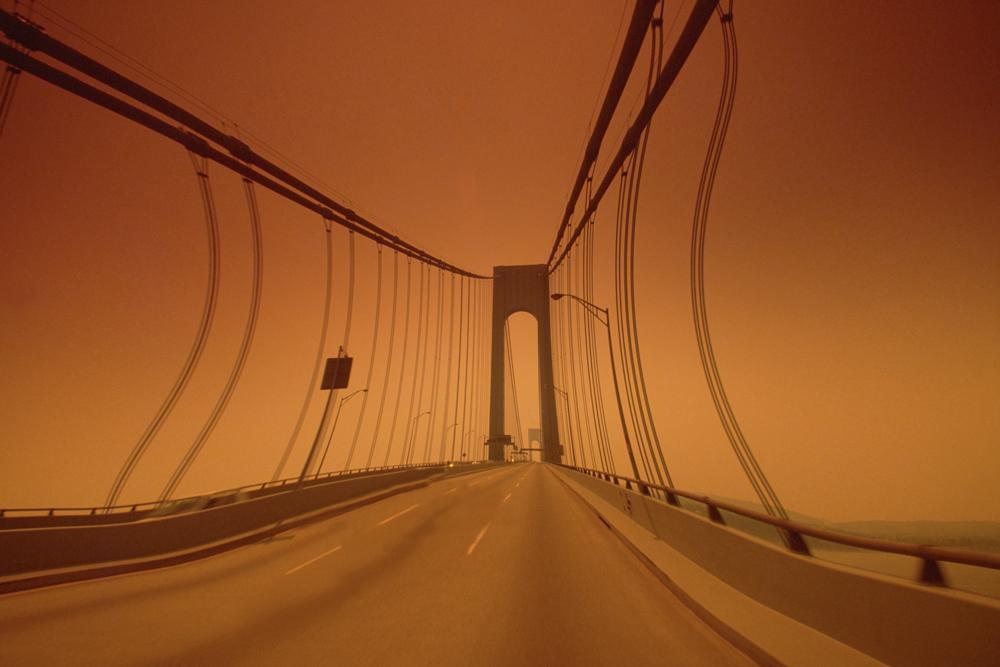}
\end{center}
   \caption{Most dissimilar images in the class of Verrazzano-Narrows bridge (images 17(left) and 19(right) in the graphical model). The image on the right is also the most isolated image in the class, based on the similarity matrix.}
\label{fig_landmark_dissim}
\end{figure}

Figure~\ref{fig_landmark_redund} shows the group of most similar images and Figure~\ref{fig_landmark_dissim} shows the most dissimilar image pair in this class, according to the SSIM measure. Analyzing the similarity matrix reveals that the right image in Figure~\ref{fig_landmark_dissim} is the most isolated image in the class.

We note that our analysis has made us familiar with images in this class and provided us with insights about their similarities and differences. We know which training images might be redundant and which image might be an anomaly. In the case of active learning, we can try to fill the gaps in training data according to the graphical model shown in Figure~\ref{fig_landmark_gm}.



%
%

\section{Conclusions and future work}  \label{sec:conclusion}

We developed a set of efficient tools for analyzing images in training sets. We showed that similar images in standard image classification datasets can be identified easy and fast, prior to training a model on them. We showed that performing this types of analysis on training and also testing sets can provide useful insights about the datasets and also the models trained on them. For example, one can quickly find redundant and influential images. By analyzing the eigen-gaps of a graph Laplacian, one can estimate the percentage of redundancies in a dataset, useful for many real world datasets.

Our method eases the computational cost barrier for analyzing the contents of image-classification datasets and therefore makes it practical for users to closely engage with the datasets and learn useful insights about their contents and their fine-level details.

Possible extension of this work is to further study the similarities and differences across training and testing sets and use that information to explain the generalization of models. Further investigating the images in relation to decision boundaries of models, during and after training, may provide useful insights on how training images shape the models by defining their decision boundaries. Moreover, our methods have applications in the context of active learning.


%

\bibliographystyle{ACM-Reference-Format}
\bibliography{refs}

\begin{acks}
R.Y. thanks Dianne O'Leary for helpful comments and advice.
\end{acks}


\end{document}